\documentclass[aps,twocolumn,prc,superscriptaddress,showpacs,nofootinbib,floatfix,amssymb,amsfonts,amsmath]{revtex4-2}

\usepackage{graphicx}
\usepackage{dcolumn}
\usepackage{bm}
\usepackage{xcolor}
\usepackage{amsmath}    
\usepackage{amsfonts}   
\usepackage{amssymb}
\usepackage{graphicx}   
\usepackage{multirow} 

\begin{document}

\title{Optimized basis of covariant density functional theory: point coupling 
functionals and excited states.}

\author{A. Dalbah}
\affiliation{Department of Physics and Astronomy, Mississippi
State University, MS 39762}

\author{A. V. Afanasjev}
\affiliation{Department of Physics and Astronomy, Mississippi
State University, MS 39762}

\author{B. Osei}
\affiliation{Department of Physics and Astronomy, Mississippi
State University, MS 39762}

\date{\today}

\begin{abstract}

     The present investigation focuses on the improvement of the accuracy of the 
description of physical observables of interest in moderately sized fermionic basis 
within the framework of covariant density functional theory. It extends previous 
study of Ref.\ \cite{OAD.25} to point coupling (PC) covariant energy density 
functionals (CEDFs) and to excited states. Using as a  benchmark  the solutions 
corresponding either to infinite fermionic basis or those extrapolated to such a 
basis  it is shown that the optimization of oscillator frequency $\hbar\omega_0$ 
of the  harmonic oscillator (HO)  basis leads to a substantial improvement in the 
description of different  physical observables  in the fermionic basis truncated at 
$N_F$.  Globally optimized scaling factors $f_{opt}(A)$ of the oscillator frequency
and the sizes $N_F^{\varepsilon}$
of the HO bases providing the required accuracy $\varepsilon$ in the calculations
of the binding energies are generated for the PC functionals. The optimization 
of the basis also significantly improves the accuracy of the description of potential 
energy curves, defining the fission barriers and fission isomers in actinides and 
superheavy nuclei, provided that the size of the basis is at least equal to $N_F=20$.
The optimization of the HO basis improves the accuracy of
the description of the energies of bound single-particle states: the only
exceptions are weakly bound neutron states with low orbital momenta $l=0$, 1 and 
2. It is demonstrated for the first time that the halo densities  of neutron halo nuclei 
generated in the coordinate space calculations are well reproduced in the calculations
with very large fermionic HO bases. 

 \end{abstract}

\maketitle

\section{Introduction}

      The basis set expansion method is a classical method of the solution of  
many quantum-mechanical  problems which is widely used in different areas of physics 
(see the introductions to Refs.\ \cite{OAD.25,CAKKMV.12} for a short review). It is 
employed in many  theoretical tools in low-energy nuclear physics where it is very 
frequently based on harmonic oscillator (HO) basis 
\cite{GRT.90,NilRag-book,K.2009,CAKKMV.12,TOAPT.24}. For example, the computer 
programs for {\it ab-initio}, shell model and density functional  theoretical calculations 
are formulated in this basis.  However, in practical applications this basis has to be
truncated because it is extremely numerically expensive to carry out the calculations
in the basis which effectively corresponds to the infinite one. This introduces numerical
errors which are difficult to quantify in the absence of the exact solutions generated
in the infinite basis.

 This is a reason why the investigations of the extrapolation features of the solutions based 
on the HO basis on the transition from small to very large (basically infinite) basis were (see Refs.\ 
\cite{CAKKMV.12,FHP.12,MEFHP.13,BEHPW.16,CK.16}) and still are (see Refs.\ \cite{MSS.25,KLMMRVW.25})
in the focus of effective field and {\it ab initio} communities. Unfortunately, such investigations 
were outside a scope of interest of density functional theoretical (DFT) community: there 
were no systematic efforts to improve or optimize the HO basis and the calculations were 
carried with oscillator length $\hbar \omega_0$ of the basis which were defined by the analysis 
of a few nuclei more than 25 years ago (see Refs.\ \cite{GRT.90,DD.97,PH.17} and the discussion 
in the introduction and Sec. VIII of Ref.\ \cite{OAD.25} for detail.).  It is only very recently that in 
Ref.\ \cite{OAD.25} a  global  optimization of the HO basis has been carried out for meson 
exchange (ME) covariant  energy density functionals (CEDFs). This optimization drastically 
increases the accuracy  of the calculations in truncated (at $N_F$) fermionic basis as compared 
with existing  procedures: here $N_F$ stands for the principal quantum number of the highest 
full fermionic shell included in the basis set expansion. Moreover, it allows to reproduce  the 
exact solutions corresponding to infinite HO basis  in  moderately sized  $N_F=20$ basis with
an accuracy of few tens of keV for binding energies at a very small fraction of computational 
cost (on average less that few \%) of the one required for exact solutions.

  The basic idea of global optimization of the HO basis is very simple and easy 
to implement into existing computer codes. It relies on the fact that scaling
factor $f$ of the oscillator length $\hbar \omega_0$ of the HO basis 
\begin{eqnarray}
\hbar \omega_0 = f \times  41 A^{-1/3} \,\, {\rm [MeV]}
\label{hbar-omega}
\end{eqnarray} 
is defined from a global comparison of the results obtained in the infinite and finite 
(truncated at $N_F$) bases. This leads to a very  high accuracy of  the calculations 
in moderately sized $N_F=20$  basis when  mass dependent oscillator frequency 
is used\footnote{Note that from 1990 the oscillator frequency of the  HO basis has 
been fixed at $\hbar \omega_0 = 41 A^{-1/3}$  [MeV] (i.e. at $f=1.0$)
in existing CDFT calculations  \cite{GRT.90,AKR.96,RGL.97,DIRHB-code.14}. 
However, this value has been defined 
from the analysis of only spherical $^{16}$O and $^{208}$Pb nuclei with the NL1 functional 
(see Ref.\ \cite{GRT.90}).}:  global rms differences   $\delta B_{rms}$  between the binding 
energies calculated in infinite and truncated bases  are only  0.025  MeV and 0.031 MeV for 
the  NL5(Z) and DD-MEZ functionals,  respectively (see Ref.\ \cite{OAD.25}).

\begin{figure}[htb]
    \centering
    \includegraphics*[width=1.0\linewidth]{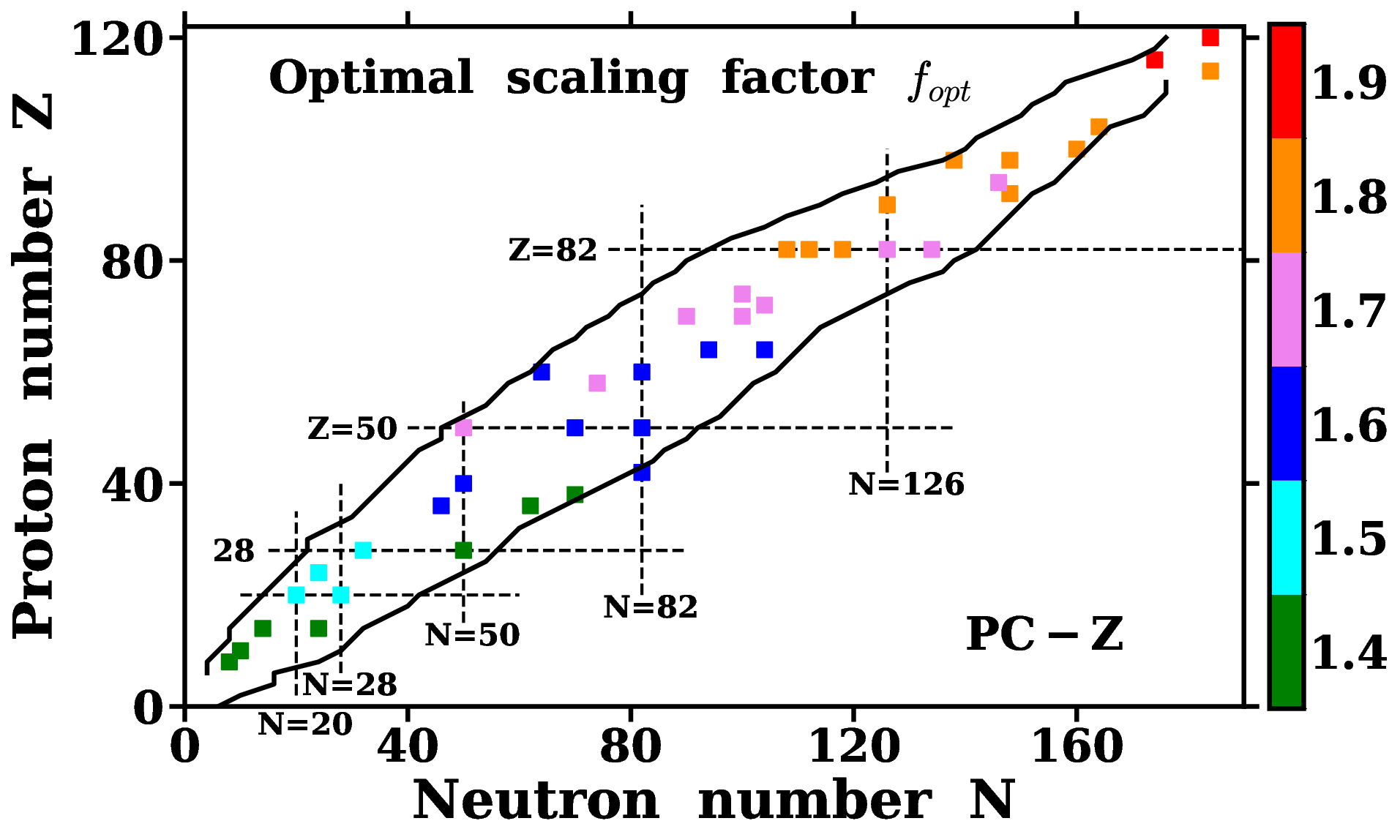}
    \caption{The nuclei (solid squares)
    analyzed in the present paper. The boundaries of experimentally known
    nuclear chart are shown by solid lines. The color of squares shows 
    the optimal values of scaling factor $f$ (see colormap).      }
    \label{map-of-nuclei}
\end{figure}

    This high accuracy of the reproduction of binding energies in moderately sized 
fermionic basis is  a consequence of a specific feature of the ME functionals in 
which the convergence to the exact solution of fermionic and mesonic energies 
proceed from above and below of the exact solution\footnote{When discussing 
the convergence to the exact solution, we consider only the monotonic part of 
the convergence curve.  According to the definitions of Sec.\ IV.A of Ref.\ 
\cite{OAD.25}, the convergence curve is monotonic in the {\it pattern A} convergence
for all values of $N_F$ and it becomes monotonic in the {\it pattern B} convergence
only for $N_F \geq N_F^{crit}$.}, respectively (see Ref.\ \cite{OAD.25}). 
However,  the convergence of total  binding energy (i.e. the sum of fermionic and 
mesonic energies) to the exact solution depends on scaling factor $f$. 
 As a consequence, an optimal  oscillator frequency $\hbar\omega_0$  of the basis 
 can be defined which provides an accurate  reproduction  of exact total binding energies 
 by the ones calculated in truncated basis at relatively low values of $N_F$.

   In contrast, the convergence of binding energies to exact solution as a function of 
the size of  basis proceeds from above with increasing $N_F$ for all values of $f$ for 
point coupling (PC) CEDFs (see Ref.\ \cite{OAD.25}). However, this was illustrated only 
for a few nuclei [spherical $^{48}$Ca and $^{208}$Pb and normal-deformed $^{240}$Pu] 
(see discussion of Figs.\ 5 and 6 in Ref.\ \cite{OAD.25} and Fig.\ 2 in 
Ref.\ \cite{NL5Z-DDMEZ-PCZ}). Thus, one of the main goals of this paper is to carry out 
systematic  investigation on how to improve the accuracy of the reproduction of infinite basis 
solutions in moderately sized fermionic basis for the PC functionals.  A large
set  of spherical and deformed even-even nuclei distributed more or less uniformly   across 
the nuclear chart is used in the present study (see Fig.\ \ref{map-of-nuclei}).  As shown 
below this leads to robust conclusions about possible ways of the optimization of the HO 
basis for the PC functionals.
 
  The second goal of the present paper is to investigate how the size of the HO basis
and the optimization of the basis affect the accuracy of the description of  excited states.  
This issue has not been covered in Ref.\ \cite{OAD.25}. Here we  consider the impact of above 
mentioned factors on potential energy curves/surfaces  which are used for the extraction 
of fission barrier heights, the energies of fission isomers  and beyond mean field effects. 
In addition, we evaluate how accurately the energies of the single-particle  states and the 
excited states based on particle-hole excitations are described in truncated fermionic basis 
as compared with infinite one and what is the impact of the optimization of the HO basis 
on the accuracy of the description of such properties.

     The third goal is to understand whether the halo structures can be analyzed in the RHB 
framework employing basis state expansion method based on HO.   At present, 
the analysis  of halo nuclei in the CDFT framework is exclusively carried out either in coordinate 
space representation (see, for  example, Refs.\ \cite{PVLR.97,ZMR.03}) or in Dirac-Woods-Saxon 
basis (see Ref.\ \cite{ZMR.03}). This is because earlier studies with the HO basis were restricted
to moderate sizes of the basis not exceeding $N_F=43$ (see, for example, Ref.\ \cite{ZMR.03}). 
As a result, they fail to reproduce neutron halo densities obtained in coordinate space 
representation. However, the  study of Ref.\ \cite{FHP.12} carried out in no-core shell model 
with chiral nucleon-nucleon interaction illustrates that such nuclei can also be  studied 
in the HO basis.

   The paper is organized as follows. 
Theoretical framework is briefly outlined in Sec.\ \ref{theory}.
Sec.\ \ref{odd-even} considers odd-even effect in the convergence of the binding
energies and its origin.  The optimization of the HO basis as a tool
for improving the convergence of point coupling CEDFs is examined
in Sec.\ \ref{Improving}.  The possibility of the description of halo nuclei in the 
HO basis and underlying physics are discussed in Sec.\ \ref{Halo}.  Sec.\ \ref{excited} 
analyses the consequences of the optimization of the HO basis for the description 
of the properties of excited states. Finally,  Sec.\ \ref{Concl} summarizes the  
results  of our paper.

\section{Brief outline of theoretical framework}
\label{theory}

      The numerical calculations are performed in the framework of relativistic 
Hartree-Boboliubov  (RHB) theory using spherical and axially deformed computer 
codes. Since technical details of such calculations are presented in Refs.\ 
\cite{NL5Z-DDMEZ-PCZ,OAD.25} we focus here only on the features which are 
relevant for the  present study. Most of the calculations in the present
paper are carried out with point-coupling  CEDF  PC-Z. The studies 
of excited states are also performed in Sec.\ \ref{excited} with the DD-MEZ 
functional since the impact of the optimization of the HO basis on such states 
has not been investigated for the ME functionals in Ref.\  \cite{OAD.25}. 
These functionals were globally optimized in Ref.\ \cite{NL5Z-DDMEZ-PCZ}.
Separable pairing interaction of Ref.\ \cite{TMR.09} with globally 
optimized strength of pairing (see Ref.\ \cite{TA.21} and  Eqs. (2) and (3) in Ref.\ 
\cite{NL5Z-DDMEZ-PCZ})  is used in the pairing channel. As recommended
in Ref.\ \cite{NL5Z-DDMEZ-PCZ}), the  $N_B=40$ bosonic basis is used in
the calculations with the DD-MEZ functional.  The numerical calculations are
typically carried out with $ngh=40$ Gauss-Hermite  and $ngl=40$ Gauss-Laguerre 
integrations points (see Sec.\ III of Ref.\ \cite{OAD.25}  for more details). However, 
for  numerical stability of solutions higher $ngh$ value is used in some spherical 
calculations with  very  large fermionic basis exceeding $N_F=70$.

\section{Odd-even effect in convergence of binding energies and its
              origin}
\label{odd-even}

\begin{figure}[htb]
    \centering
    \includegraphics*[width=1.0\linewidth]{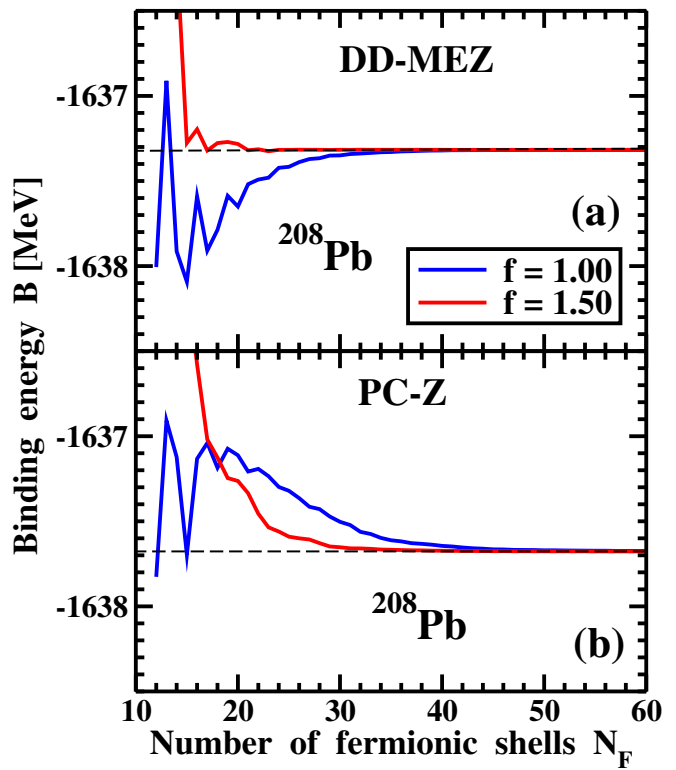}
    \caption{The binding energies $B$ of the ground state of  spherical 
    $^{208}$Pb nucleus  as a function of $N_F$ for scaling factor $f=1.0$ and
    $f=1.5$. The calculations have been carried out in step of $\Delta N_F=1$         
    for the DD-MEZ and NL5(Z) functionals. Thin dashed line shows the 
    exact value of binding energy corresponding to infinite basis. 
    }
    \label{OE-effect-208Pb}
\end{figure}

   In the CDFT framework, the convergence of binding energies as a function 
of the size of the basis has always been investigated in step of $\Delta N_F=2$ 
(see, for example, the discussion in Sec.\ V of Ref.\ \cite{TOAPT.24} 
and Ref.\ \cite{ZMR.03}). However, the detailed analysis of all classes of CEDFs
reveals odd-even staggering in binding energies of the convergence curve
which is seen when the calculations are performed is step of  $\Delta N_F=1$ 
(see Fig.\ \ref{OE-effect-208Pb}). This effect is most pronounced at low 
$N_F=10-20$  values  and its magnitude gradually decreases with increasing 
$N_F$ so that it disappears at very large $N_F$.  Note that the removing either all odd or all even 
values of $N_F$  from consideration makes a convergence curve smoother without 
substantial fluctuations. This is the most likely reason why in the past the convergence of
binding energies as a function of the size of fermionic basis has been studied only in 
step of $\Delta N_F=2$. 

\begin{figure}[htb]
    \centering
    \includegraphics*[width=1.0\linewidth]{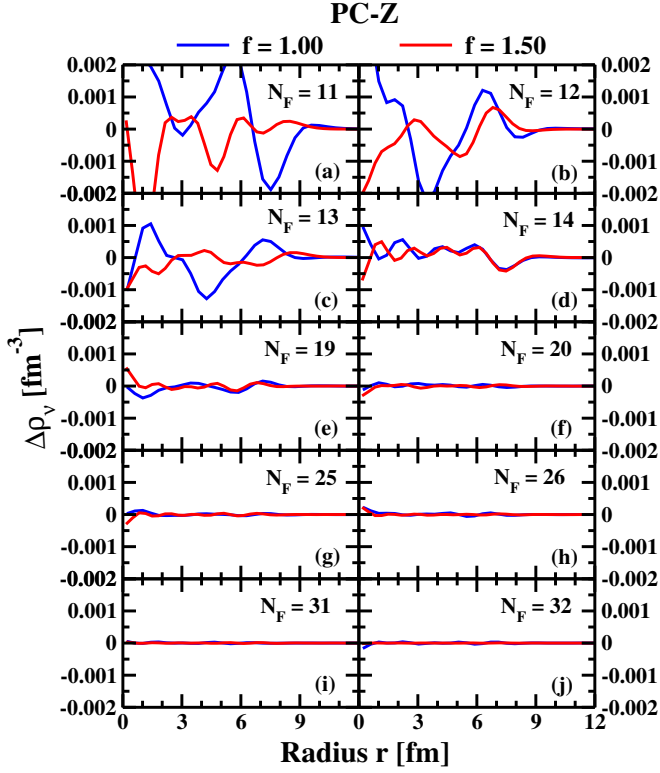}
    \caption{The differences $\Delta \rho_{\nu}[N_F](r) = \rho_{\nu}[N_F](r) - \rho_{\nu}[N_F-1](r)$    
of calculated neutron densities of $^{208}$Pb obtained at the Gauss-Hermite  
integration  points at indicated values of $N_F$. The results are presented for the
$f=1.0$ and $f=1.5$ values of scaling factor. The panels (a)-(d) show the results
for sequential filling of the $N_F=11$, 12, 13 and 14 shells. The effect becomes
smaller at higher $N_F$ values. Thus, in other panels the impact of filling of
odd and even $N_F$ shells in selected pairs of the shells (19 and 20, 25 and 26 as 
well as 31 and  32) is shown in order to illustrate that odd-even effect in density
changes is present even for higher $N_F$ values. 
    \label{OE-density}
    }
\end{figure}

    To our knowledge odd-even staggering in binding energies of the convergence
curve has never been studied in the DFT framework. Thus, it is important to understand
the origin of this effect. To do that
the differences $\Delta \rho_{\nu}[N_F](r)$ of neutron densities in $^{208}$Pb calculated 
at $N_F$ and $N_F-1$ are shown in Fig.\ \ref{OE-density}. One can see that these 
differences  are large at low values of $N_F$, but they decrease with increasing $N_F$ 
and they become very small at large values of $N_F$. In addition, the distribution of  
$\Delta \rho_{\nu}[N_F](r)$ as a function of radial coordinate changes with increasing 
$N_F$. 

    Let us discuss these results in more detail for the scaling factor $f=1.0$. One can 
see that the addition of the $N_F=11$ shell increases neutron density in the central 
region of the nucleus and at $r\approx 5.5$ fm, but decreases it at $r\approx 7.5$ fm 
(see blue line in Fig.\ \ref{OE-density}(a)). In contrast, the addition of the $N_F=12$ 
shell leads to the increase of neutron density in the central region and at $r\approx 6.5$ 
fm and to its decrease at $r\approx 3.5$ fm (see blue line in Fig.\ \ref{OE-density}(b)).
The addition of the $N_F=13$ shell leads to an increase of neutron density at $r\approx 1.5$ 
fm and at $r\approx 7$ fm but to its decrease in the central region of the nucleus
and at $r\approx 4.3$ fm. The underlying nodal structure of the wave functions and 
single-particle densities,  self-consistency and proton-neutron interaction effects on the 
single-particle densities  (see Refs.\ \cite{PA.22,PA.23}) are responsible for this non-regular 
behavior of $\Delta \rho_{\nu}[N_F](r)$ with sequential addition of the shells. These non-regular 
changes in the densities with  increasing  $N_F$ cause self-consistent feedback to binding 
energies which leads to some non-regular fluctuations of these energies around
some smooth trend with increasing $N_F$, i.e. odd-even staggering in convergence
curves of binding energies.

    As illustrated in Fig.\ \ref{OE-effect-208Pb}, this effect strongly depends on scaling 
factor $f$ of oscillator frequency $\hbar \omega_0$ of the basis. One can see that the 
increase of $f$ from 1.0 to 1.5 suppresses substantially this effect  and leads to a 
smoother convergence curves both for the PC and DDME functionals. This is due to 
the fact that at a given $N_F$  on average the magnitude of the $\Delta \rho_{\nu}[N_F](r)$
values are smaller for the $f=1.5$ calculations as compared with the $f=1.0$ ones
(see Fig.\ \ref{OE-density}).

\begin{figure}[htb]
    \centering
    \includegraphics*[width=1.0\linewidth]{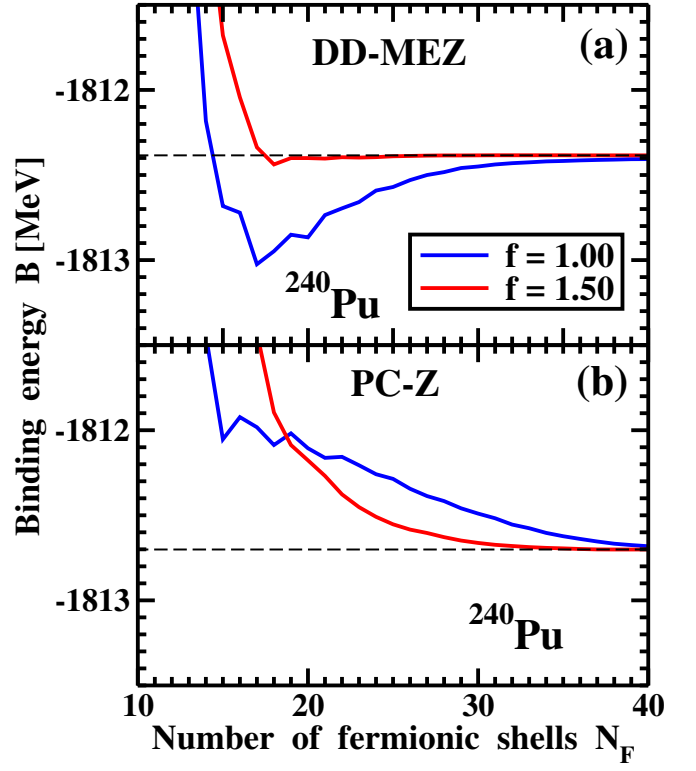}
    \caption{The same as Fig. \ref{OE-effect-208Pb} but for normal-deformed
    ground state of $^{240}$Pu. Thin dashed line shows the value 
    of binding energy extrapolated to infinite fermionic basis.    
    }
    \label{OE-effect-240Pu}
\end{figure}

    The deformation of the nucleus reduces the magnitude of odd-even staggering due 
to additional mixing of the single-particle states caused by the deformation and pairing 
(compare blue lines in Figs.\ \ref{OE-effect-208Pb} and \ref{OE-effect-240Pu}). The 
increase of scaling factor $f$ from 1.0 to 1.5 leads to a substantial suppression of this
staggering (see Fig.\ \ref{OE-effect-240Pu}).

\begin{figure}[htb]
    \centering
    \includegraphics*[width=.65\linewidth]{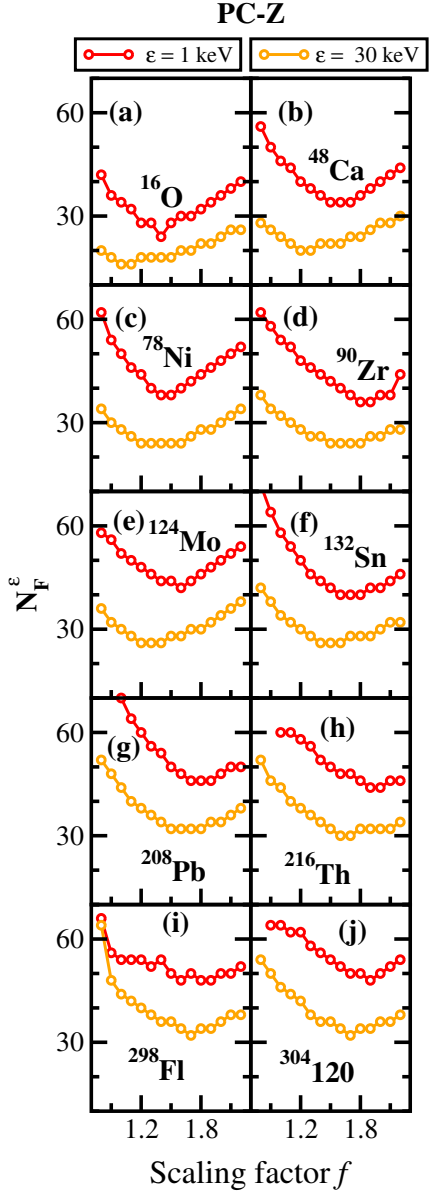}
    \caption{The values of $N_F^{\varepsilon}$ at which the calculations
    in the $N_F^{\varepsilon}$ basis reproduce infinite basis solutions with an accuracy 
    better than $\varepsilon$,  i.e. $|B(N_F^{\varepsilon}) - B(N_F=\infty)| \leq \varepsilon$. 
    The results are presented for $\varepsilon=1$ keV and $\varepsilon=30$ keV.
                  }
    \label{fig:spherical_pcz}
\end{figure}

   This odd-even staggering of binding energies in the convergence curves is 
expected to be active also in non-relativistic DFTs based on the Skyrme and
Gogny functionals because the underlying physical mechanism of the connection
between the density changes and modifications of binding energies is similar. 
However, this effect is not expected to play a role in the models based on 
phenomenological potentials such as Woods-Saxon or Nilsson since 
feedback loop between densities and binding energies (i.e. self-consistency 
effects) is not active in them: the changes in proton and neutron densities 
do not affect these potentials. Consequently, in such potentials the convergence 
curves of  binding energies as a function of $N_F$ are expected to behave 
smoother  with better extrapolation properties to infinite basis.
 
\section{Improving the convergence of point coupling covariant energy 
              density functionals}
\label{Improving}

\subsection{Spherical nuclei}

     Full convergence of binding energies to the  exact solution corresponding to infinite basis can be 
obtained for all nuclei of interest in spherical RHB code. This is clear advantage 
of spherical nuclei since exact and truncated solutions can be compared without
any extrapolations. The values of $N_F^{\varepsilon = 1\,{\rm keV}}$ at which the full convergence
within $\varepsilon = 1$ keV error is 
reached and their dependence on scaling factor $f$ are shown in Fig.\ 
\ref{fig:spherical_pcz}.  Most of the nuclei displayed on this figure are spherical in 
the ground state, but for some  of them (such as $^{216}$Th and $^{298}$Fl) the 
calculations are restricted for spherical shape.

    For each nucleus there is a value of scaling factor $f$ which leads to the minimum 
in the $N_F^{\varepsilon = 1\,{\rm keV}}$ curves.  For example, these are the $f$ values 
of 1.4, 1.4, and 1.6 for the $^{16}$O, $^{78}$Ni and $^{124}$Mo nuclei, respectively [see 
Figs.\ \ref{fig:spherical_pcz}(a), (c) and (e)].  However, for some nuclei such as 
$^{208}$Pb, $^{298}$Fl and $^{304}$120 the minimum of the $N_F^{\varepsilon = 1\,{\rm keV}}$ curves 
represents a plateau formed by several values of $f$ [see Figs.\ \ref{fig:spherical_pcz}(g), 
(i) and (j)]. The gain in convergence due to the optimization of scaling factor $f$ is especially 
pronounced in light nuclei: for example, the transition from $f=0.8$ to $f=1.5$ reduces 
the HO basis required for a full convergence by almost 20 fermionic shells [see Figs.\ 
\ref{fig:spherical_pcz}(b)]. In contrast, this gain is weaker in superheavy nuclei: the same 
transition from $f=0.8$ to $f=1.5$ in $^{298}$Fl  leads to a reduction of only approximately 
ten shells. 

\begin{figure}[htb]
    \centering
    \includegraphics*[width=.6\linewidth]{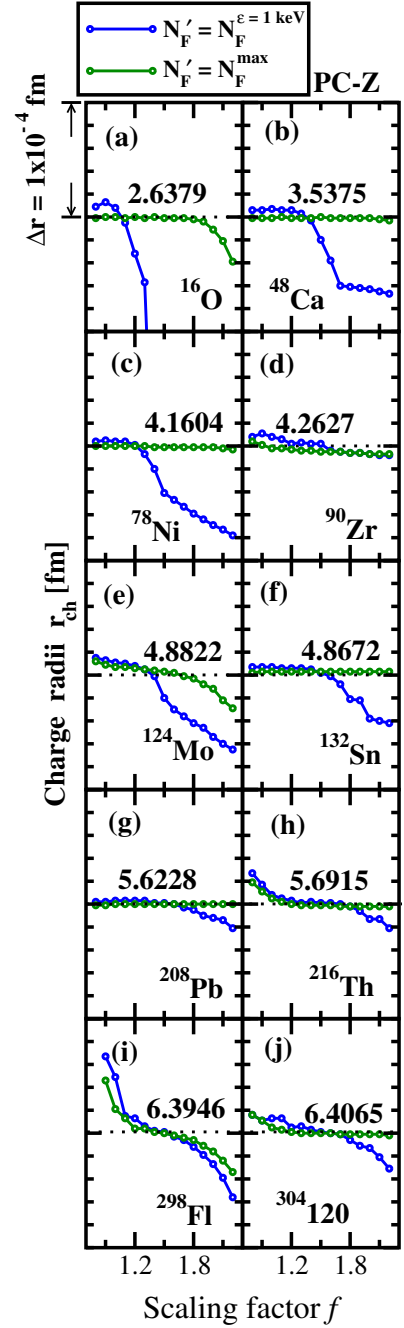}
    \caption{The calculated values of charge radii $r_{ch}$ as a function of 
                 scaling factor $f$ for indicated spherical nuclei. Dotted line shows the value
                 of $r_{ch}$ corresponding to approximate value of plateau of the $r_{ch}(f)$ 
                 function; its numerical value is displayed.  The calculations are carried out 
                 with the PC-Z functional. See text for further detail.}
    \label{fig:rms}
\end{figure}

   To minimize numerical cost the CDFT calculations  are carried out  in the basis truncated at $N_F$.
This approach is sound only in the case when numerical  errors of the truncation are well established. 
However, this can be achieved only when either exact  solution is known (as in the case of spherical nuclei) 
or  reliable extrapolation procedure  to infinite fermionic basis can be employed (as in the case of deformed 
nuclei). The knowledge of exact or extrapolated solutions allows to evaluate the size of the truncated HO 
basis $N_F^{\varepsilon}$ required for numerical calculations with a given numerical error $\varepsilon$. In 
most of the applications the numerical error at the level of $\varepsilon= 30$ keV would be more than 
sufficient. For such error the calculations show a drastic reduction (by almost 20 fermionic shells) of the 
size of the HO basis (compare $N_F^{\varepsilon = 30\,{\rm keV}}$ and $N_F^{\varepsilon = 1\,{\rm keV}}$ 
curves in Fig.\ \ref{fig:spherical_pcz}). Note also  that the truncated solution is always less bound (by 
approximately $\varepsilon$) than the exact one since the binding energies converge to the exact one 
from above for the PC functionals.

    Note that $N_F^{\varepsilon = 1\,{\rm keV}}$ and $N_F^{\varepsilon = 30\,{\rm keV}}$ curves calculated 
 with the PC-PK1 \cite{PC-PK1},  DD-PC1 \cite{DD-PC1}, and PC-Y \cite{NL5Z-DDMEZ-PCZ} functionals 
 are very similar to those shown in Fig.\ \ref{fig:spherical_pcz} which are obtained with the PC-Z one: the 
 local deviations from the PC-Z results at specific values of $f$ rarely exceed 4.  This clearly indicates that the 
convergence properties of binding energies are similar for the CEDFs belonging to the PC class 
of the functionals. This feature was seen earlier for the DDME and NLME functionals in Ref.\ 
\cite{OAD.25}.

     The variations of charge radii $r_{ch}$ as a function of scaling factor $f$ are shown in
Fig.\ \ref{fig:rms}. The blue curves with circles in this figure show charge radii calculated 
at the values of $N_F'(f)= N_F^{\varepsilon = 1\,{\rm keV}}(f)$ provided by red curves in 
Fig.\ \ref{fig:spherical_pcz}.  As discussed in Ref.\ \cite{OAD.25} the scaling 
factors $f$ significantly smaller than 1.0 are not recommended because of the fast variation 
of binding energies with the  change of $f$ in truncated bases.   The analysis of this paper 
also indicates that the exact results are best reproduced in truncated basis if the value of $f$ 
is located between $\approx 1.0$ and $\approx 2.2$.  Figure \ref{fig:rms} shows that if one 
restricts the range of the change of $f$ to $f=1.0 - 2.0$, then the variation of $r_{ch}$  calculated 
with $N_F'(f)= N_F^{\varepsilon = 1\,{\rm keV}}(f)$ is typically below $8\times 10^{-5}$ fm. 
The only exception from this rule are the $^{16}$O and $^{298}$Fl nuclei [see 
\ref{fig:spherical_pcz}(a) and (i)].

  These results show that the convergence of charge radii as a function
of $N_F$ is slower than the one for binding energies. Indeed, the accuracy of the description 
of binding energies at the level of 1 keV reached in the  $N_F^{\varepsilon = 1\,{\rm keV}}(f)$ 
calculations means that they are described globally with an accuracy of approximately 
10$^{-7}B$. However, the results presented in Fig.\ \ref{fig:spherical_pcz} 
shown that for the same truncation of the basis the charge radii are described with an accuracy 
of approximately $10^{-6}r_{ch}$\footnote{This estimate is obtained assuming 
average charge radius of the nuclei in known nuclear chart at the level of 5 fm and average
variation of $r_{ch}$ for the $f=1.0 - 2.0$ range at the level of $\approx 4\times 10^{-5}$ fm
(see Fig.\ \ref{fig:spherical_pcz}).}.  This is consistent with the results of no-core shell model
calculations formulated in the HO basis which show slower convergence for charge radii as 
compared with that for binding energies (see, for example, Refs.\ \cite{MSS.25}).  Note 
that above mentioned variations 
of $r_{ch}$ as a function of $f$ are not critical since they are significantly smaller than the accuracy 
of the  measurements of absolute values of charge radii (see Ref.\  \cite{AM.13}). 

    It turns out that further increase of the basis considerably suppresses the dependence of 
calculated charge radii on scaling factor $f$. This is shown by green solid lines with open
circles in Fig.\ \ref{fig:spherical_pcz}  which were calculated in the maximum $N_F^{max}$ 
fermionic basis numerically achievable at a given value of scaling factor $f$.  Note that
with an exception of $^{298}$Fl,  $N_F^{max}(f)$  is larger  than $N_F^{\varepsilon = 1\,{\rm keV}}(f)$ 
shown in Fig.\ \ref{fig:spherical_pcz} by a factor of at least 20.

\subsection{Deformed nuclei}
  
   The analysis of the convergence of binding energies in deformed nuclei is 
complicated by the fact that numerical calculations in axially deformed RHB code 
are possible only in the fermionic bases with $N_F$ up to 40 (see Ref.\ 
\cite{NL5Z-DDMEZ-PCZ}).  Since the convergence to exact solution in spherical 
and deformed nuclei of a given region of nuclear chart takes place at similar 
$N_F$ values, the results presented in Fig.\ \ref{fig:spherical_pcz} clearly 
indicate that in the majority of deformed nuclei the exact solution for the PC 
functionals cannot be defined by direct numerical calculations.

\begin{figure}[htb]
    \centering
    \includegraphics*[width=0.9\linewidth]{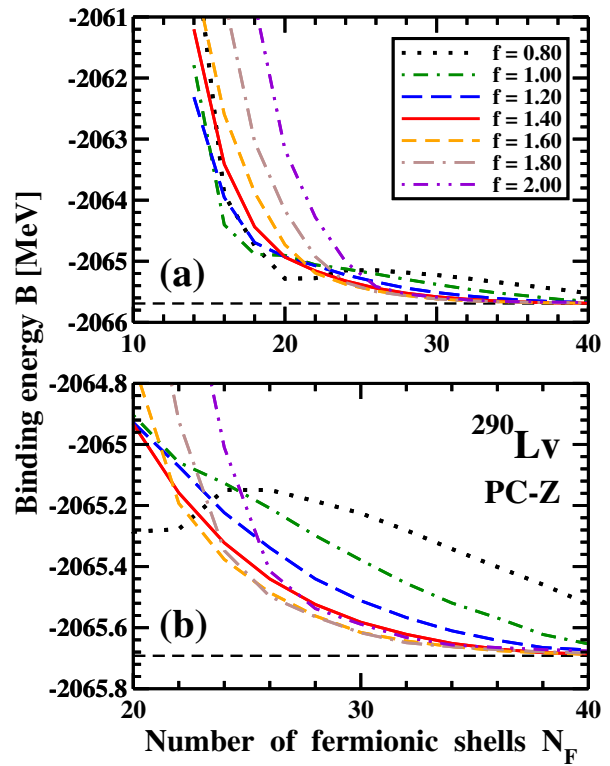}
    \caption{The binding energies of the superdeformed $(\beta_2 \approx 0.55)$ minimum 
     of the $^{290}$Lv nucleus as a function of $N_F$ for different values of
     scaling factor $f$.   Panel (b) shows the results in enhanced scale.  Thin dashed 
     line shows the extrapolated value of binding energy corresponding to infinite basis. 
    \label{Lv290-conv}
    }
\end{figure}

     Thus, some extrapolation procedure to infinite basis has to be employed
to obtain extrapolated value of binding energy. This procedure based on the 
analysis of the rate of change of binding energy as a function of $N_F$ is
a subject of discussed below constraints. 

  First, it has to based on the monotonic
part of the convergence curve. As discussed in Sect.\ IV of Ref.\ \cite{OAD.25}
on the example of $^{208}$Pb and $^{240}$Pu nuclei (see Figs. 4 and 6 in 
this paper), the convergence curves behave monotonically only above some
critical value of $N_F^{crit}$ which depends on the nucleus, scaling factor $f$ 
and the functional. The systematic analysis of convergence curves of the 
nuclei studied in the present paper reveals that non-monotonic behavior similar 
to that seen in Fig.\ 6 of Ref.\ \cite{OAD.25} and in the $f=0.8$ case of Fig.\ 
\ref{Lv290-conv}  is a general feature of the PC functionals in the calculations 
with low values of $f=0.8, 1.0$, and 1.2. The same analysis reveals that the
convergence curves behave monotonically for higher values of $f$ (see, for 
example, Fig.\ \ref{Lv290-conv}).  However, these curves are strongly 
down-sloping at low and moderate values of $N_F$.  As a result, the
extrapolations to infinite $N_F$ values based on the data from this branch of the 
convergence curve will either provide no convergence or the convergence to a 
wrong extrapolated value.

\begin{figure}[htb]
    \centering
    \includegraphics*[width=.9\linewidth]{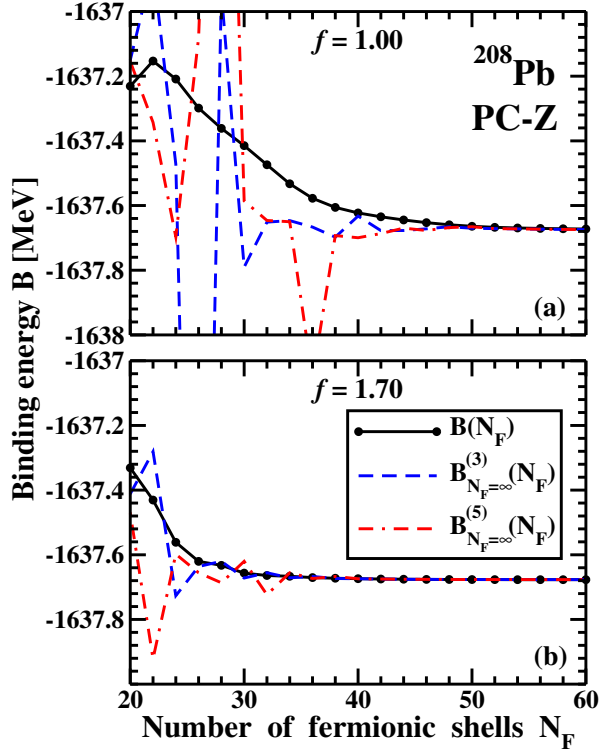}
    \caption{The comparison of calculated values of binding energies $B(N_F)$
                  and those [i.e. $B^{(3)}_{N_F=\infty}(N_F)$ and 
                  $B^{(5)}_{N_F=\infty}(N_F)$, see Eqs.\ (\ref{3-point}) and (\ref{5-point})]                                   
                  obtained at $N_F$ via the extrapolations to
                  infinite basis using Shanks transformations. Panels (a) and (b) show the
                  results for scaling factors $f=1.0$ and 1.7, respectively.                 
    \label{fig-Shanks}
    }
\end{figure}

     Second, for a reliable extrapolation of binding energies to infinite basis the 
convergence curves should also be smooth. However, because of self-consistency 
between the changes in the densities and binding energies with increasing $N_F$
discussed in Sec.\ \ref{odd-even} this is difficult to achieve for the PC functionals when 
the extrapolations are performed at low or moderate values of $N_F$.  To account 
for the variations of  binding energies with increasing $N_F$  the 
extrapolations have to include binding energies calculated for at least  three values 
of  $N_F$. Because of odd-even effect in convergence of  binding energies (see 
Sect.\ \ref{odd-even}), these extrapolations cannot be based on the combinations
of odd and even values of $N_F$. However, even if one type of the $N_F$ values (odd
or even) is considered, the above mentioned self-consistency effects substantially 
reduce the reliability of the extrapolation procedures at low and medium values of
$N_F$. This is illustrated in Fig.\ \ref{fig-Shanks}  which compares the extrapolated values 
of binding energies $B^{(3)}_{N_F=\infty}(N_F)$ and $B^{(5)}_{N_F=\infty}(N_F)$ obtained 
at given values of  $N_F$ with 3- and 5-point Shanks transformations (see Refs.\ 
\cite{Shanks.55,Wynn.56})
\begin{eqnarray} 
B^{(3)}_{N_F=\infty}(N_F) = \frac{B(N_F+2)B(N_F-2) - (B(N_F))^2}{B(N_F+2)-2 B(N_F)+B(N_F-2)}, 
\nonumber \\ 
\label{3-point} \\
B^{(5)}_{N_F=\infty}(N_F) = \qquad \qquad \qquad \qquad \qquad \qquad \qquad \nonumber \\ 
= \frac{B^{(3)}_{N_F=\infty}(N_F-2) B^{(3)}_{N_F=\infty}(N_F+2) - (B^{(3)}_{N_F=\infty}(N_F))^2}
{B^{(3)}_{N_F=\infty}(N_F+2) - 2B^{(3)}_{N_F=\infty}(N_F) + B^{(3)}_{N_F=\infty}(N_F-2)}.
 \nonumber  \label{5-point}
 \\
\end{eqnarray} 
One can see that in the calculations with $f=1.0$ the extrapolation errors exceed or are in the
vicinity of 100 keV for basis sizes $N_F^{\varepsilon=100\,\,{\rm keV}}  \leq 40$ [see Fig.\ \ref{fig-Shanks}(a)]. The use of
larger scaling factor $f=1.7$ substantially improves the convergence and extrapolation
errors become smaller than 40 keV at $N_F^{\varepsilon=40\,\,{\rm keV}}\approx 30$ [see Fig.\ \ref{fig-Shanks}(b)]. 
Note that the use of 5-point Shanks transformation does not offer any benefits as 
compared with the 3-point one.  We also employed Richardson extrapolation of Ref.\ 
\cite{RK.93} to  obtain extrapolated binding energies: similar results to those generated
by the 3-point  Shanks transformation have been obtained. 

\begin{figure}[htb]
    \centering
    \includegraphics*[width=0.99\linewidth]{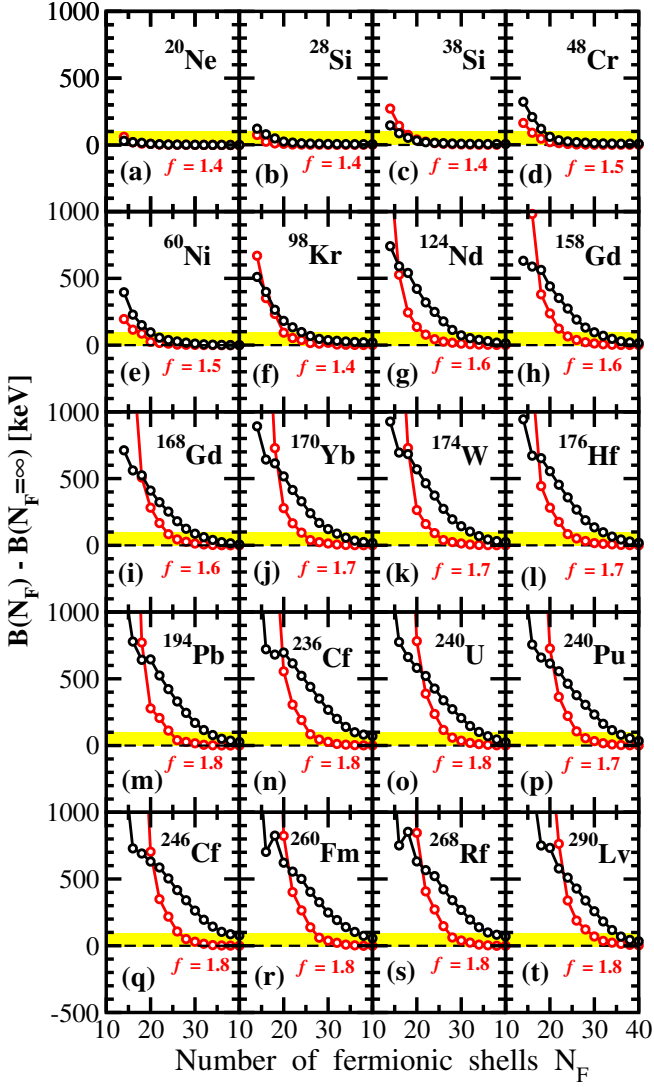}
    \caption{ The convergence curves 
    $B(N_F)-B(N_F=\infty)$ for $f=1.00$ (black curves) and $f_{opt}$ 
    (red curves).
    The optimal values of $f_{opt}$ are shown in red. Yellow shading shows 100 keV 
    error with respect of infinite basis solutions.
    \label{deformed-conv-summary}
    }
\end{figure}

\begin{table}[htb]
\centering
\caption{The number $N_F^{\varepsilon}$ of fermionic shells required to reach 
the accuracy $\varepsilon$ of the reproduction of infinite basis solution (see columns 3 and 
4) in the nuclei under study. Column 2 shows the values of optimal scaling factor $f_{opt}$ 
for which the calculations have been carried out. The equilibrium quadrupole  deformations 
$\beta_2$ of the solutions are displayed in column 5. 
\label{Table-opt}
}
\begin{tabular}{c c c c c}
\hline \hline
Nucleus & $f_{opt}$ & \,\,$\varepsilon = 0.1$ MeV\,\, & $\varepsilon = 0.03$ MeV\,\, & $\beta_2$ \\ \hline
    1              & 2    &  3  & 4   & 5  \\ \hline
$^{16}$O     & 1.4 & 12 & 16 & 0.00 \\
$^{20}$Ne    & 1.4 & 14 & 16 & $+0.28$ \\
$^{28}$Si    & 1.4 & 14 & 16 & $-0.34$ \\
$^{38}$Si    & 1.4 & 18 & 22 & $+0.32$ \\
$^{40}$Ca    & 1.5 & 12 & 18 & 0.00 \\
$^{48}$Ca    & 1.5 & 16 & 20 & 0.00 \\
$^{48}$Cr    & 1.5 & 16 & 20 & $+0.25$ \\
$^{60}$Ni    & 1.5 & 18 & 20 & $+0.17$ \\
$^{78}$Ni    & 1.4 & 20 & 22 & 0.00 \\
$^{82}$Kr    & 1.6 & 20 & 22 & $+0.05$ \\
$^{90}$Zr    & 1.6 & 20 & 22 & 0.00 \\
$^{98}$Kr    & 1.4 & 20 & 26 & $-0.22$ \\
$^{100}$Sn   & 1.7 & 20 & 22 & 0.00 \\
$^{108}$Sr   & 1.4 & 22 & 26 & 0.00 \\
$^{120}$Sn   & 1.6 & 22 & 26 & 0.00 \\
$^{124}$Mo   & 1.6 & 22 & 28 & 0.00 \\
$^{124}$Nd   & 1.6 & 22 & 26 & $+0.40$ \\
$^{132}$Ce   & 1.7 & 22 & 26 & $+0.19$ \\
$^{132}$Sn   & 1.6 & 22 & 26 & 0.00 \\
$^{142}$Nd   & 1.6 & 22 & 26 & 0.00 \\
$^{158}$Gd   & 1.6 & 24 & 28 & $+0.34$ \\
$^{160}$Yb   & 1.7 & 24 & 28 & $+0.21$ \\
$^{168}$Gd   & 1.6 & 24 & 28 & $+0.35$ \\
$^{170}$Yb   & 1.7 & 24 & 28 & $+0.35$ \\
$^{174}$W    & 1.7 & 24 & 28 & $+0.34$ \\
$^{176}$Hf   & 1.7 & 24 & 28 & $+0.31$ \\
$^{190}$Pb  & 1.8 & 24 & 28 & $-0.17$ \\
$^{194}$Pb   & 1.8 & 24 & 28 & $-0.16$ \\
$^{200}$Pb   & 1.8 & 24 & 28 & 0.00 \\
$^{208}$Pb   & 1.7 & 26 & 30 & 0.00 \\
$^{216}$Pb   & 1.7 & 26 & 30 & 0.00 \\
$^{216}$Th   & 1.8 & 26  & 30 & 0.00 \\
$^{236}$Cf   & 1.8 & 26 & 30 & $+0.26$ \\
$^{240}$Pu   & 1.7 & 26 & 30 & $+0.29$ \\
$^{240}$U    & 1.8 & 28 & 32 & $+0.29$ \\
$^{246}$Cf   & 1.8 & 26 & 30 & $+0.30$ \\
$^{260}$Fm   & 1.8 & 28 & 32 & $+0.27$ \\
$^{268}$Rf   & 1.8 & 28 & 32 & $+0.26$ \\
$^{290}$Lv   & 1.8 & 30 & 34 & $+0.54$ \\
$^{298}$Fl   & 1.8 & 28 & 34 & 0.00 \\
$^{304}$120  & 1.9 & 30 & 34 & 0.00 \\
\hline \hline
\end{tabular}
\end{table}

   It is interesting that the $B(N_F)$ curves are smoother than the $B^{(3)}_{N_F=\infty}(N_F)$ 
and $B^{(5)}_{N_F=\infty}(N_F)$ ones which fluctuate substantially up to above mentioned
 $N_F^{\varepsilon}$ values. Note that for the $N_F > N_F^{\varepsilon}$  basis sizes these 
curves are very close to each other or even overlap. This suggests the following approach for 
the improvement of the accuracy of the calculations of binding energies with controllable errors 
for the PC functionals:
\begin{itemize} 
\item 
Define optimized scaling factor $f_{opt}$ of oscillator frequency for a given nucleus in such 
a way  that nearly flat (as a function of $N_F$) part of the convergence curve is reached at 
the lowest value of $N_F$ as compared with the calculations employing other values of $f$. 
For  example, in the case of $^{290}$Lv, the binding energies obtained with $f=1.8$ becomes 
the lowest among considered cases at $N_F=26$ (see Fig.\ \ref{Lv290-conv}).
 
\item
 For a subset of deformed nuclei shown in Fig.\ \ref{map-of-nuclei} define the infinite 
basis  solutions by  extrapolating  the results calculated with $N_F=36, 38$ and 40 
and $f_{opt}$ to $N_F=\infty$.  Then determine the value of $N^{\varepsilon}_F$ at which 
the difference $B(N_F^{\varepsilon})-B(N_F=\infty)$ does not exceed desired error 
$\varepsilon$ in the calculations of binding energies.

\item
  Define the global trends for $N^{err}_F$ and $f_{opt}$  based on the calculations of spherical 
and deformed nuclei. This is in line with the results obtained in Ref.\ \cite{OAD.25} for 
meson exchange functionals which show reasonably smooth variation of $f_{opt}$ with 
mass number $A$.

\end{itemize}

This approach is illustrated for deformed nuclei in Fig.\ \ref{deformed-conv-summary}
which compares convergence curves calculated with $f=1.0$ and $f_{opt}$. One can see 
that the value of  $f_{opt}$ increases from 1.4 up to 1.8  with increasing  mass of 
nucleus. In all nuclei, the convergence to $B(N_F=\infty)$ proceeds substantially 
faster in the $f_{opt}$ basis as compared with the $f=1.0$ one. Note that
the extrapolation to infinite basis is impossible in the $f=1.0$ calculations
of deformed actinides and superheavy nuclei  restricted\footnote{The
$N_F=40$ is the maximum size of the basis in which the RHB calculations are
possible with separable pairing (see Ref.\ \cite{NL5Z-DDMEZ-PCZ}).} to $N_F=40$ since the respective 
convergence curves decrease nearly linearly as a function of $N_F$ in the $N_F=20-40$ 
range [see Figs.\ \ref{deformed-conv-summary}(n),(o),(p),(q),(r),(s) and (t)]. This is  a reason 
why such nuclei were excluded from the global fits of the PC functionals in Refs.\ 
\cite{TOAPT.24,NL5Z-DDMEZ-PCZ}.  In contrast, such extrapolations are straightforward 
in the calculations with $f_{opt}$.  They also allow to evaluate how far are the $(N_F=40,
f=1.0)$ solutions from the ones obtained in the infinite basis [see Figs.\ 
\ref{deformed-conv-summary}(n),(o),(p),(q),(r),(s) and (t)].

\begin{figure}[htb]
    \centering
     \includegraphics*[width=1.00\linewidth]{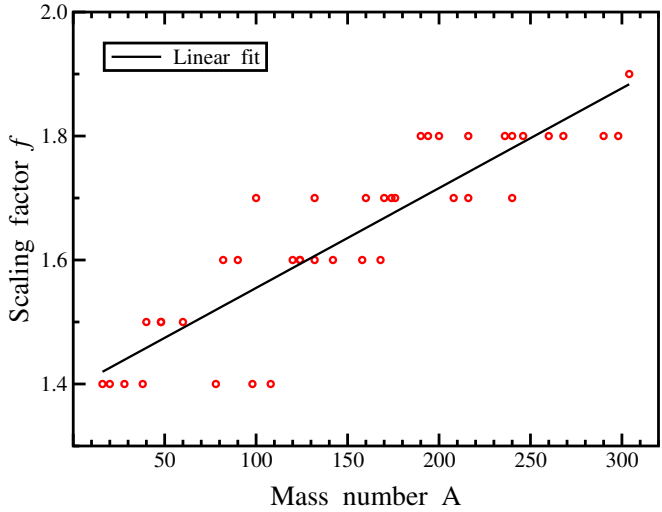}
   \caption{The distribution of optimal scaling factors $f_{opt}$ 
    as a function of  mass number $A$. Based on the results presented 
    in Table \ref{Table-opt}.   
    \label{evol-NF}
    }
\end{figure}

   Table \ref{Table-opt} summarizes optimal scaling factors $f_{opt}$ and the basis sizes 
$N_F^{\varepsilon}$ required to reach the $\varepsilon$ accuracy in the description of binding 
energies.  Note that this table combines the results for all spherical and deformed nuclei 
displayed in Fig.\ \ref{map-of-nuclei} and presents them for the  $\varepsilon=0.1$ MeV and 
$\varepsilon=0.03$ MeV errors. One can see 
that the difference between $N_F^{\varepsilon}$ values obtained for spherical and deformed nuclei in 
a given part of nuclear chart  is modest indicating that the deformation of the nuclei does not 
affect significantly  the  pattern of the convergence of binding energies.  This table also reveals
that the size of basis $N_F^{\varepsilon}$ gradually increases with increasing mass number $A$.

    Fig.\ \ref{evol-NF} shows the distribution of optimal scaling factors $f_{opt}$ as a function
of mass number $A$. One can see that there is no pronounced mass dependence of these
factors reflecting that above some $N_F$ value the solutions with a number of the $f$ values come
very close to the solution with  the $f_{opt}$ one. This is clearly seen, for example, in $^{290}$Lv where
the $f=1.60$ solution comes extremely close to the optimal solution with $f=1.8$ closely 
followed by the $f=1.4$ and $2.0$ solutions (see Fig.\ \ref{Lv290-conv}). The systematics
of spherical solutions presented in Fig.\ \ref{fig:spherical_pcz} also show weak dependence 
of $N_F^{\varepsilon}$ on the $f$ value in some range of the $f$ values.  Keeping this
in mind one can suggest an approximate expression for optimal scaling factor 
\begin{eqnarray}
f_{opt}(A) \approx 1.394 + 0.00161 A
\label{f-mass-dep}
\end{eqnarray}
which can be used in global calculations and for nuclei in which above mentioned 
optimization of $f$ has not been carried out.

 In contrast, the minimum fermionic basis $N_F^{\varepsilon}$ providing the accuracy of 
the description of binding energies at the $\varepsilon$ level shows much more 
pronounced mass dependence (see Fig.\ \ref{evol-NF})  and it can be reasonably well 
parametrized via 
\begin{eqnarray}
N_F^{\varepsilon = 30\,{\rm keV}}(A)  \approx 17.45 + 0.05780 A, \\
N_F^{\varepsilon= 100\,{\rm keV}}(A)  \approx 14.22 + 0.05368 A.
\end{eqnarray}
   It is interesting to compare these results with the ones obtained for meson 
exchange functionals in Ref.\ \cite{OAD.25} which showed that one can reach 30 keV 
global accuracy of the reproduction of binding energies calculated in infinite fermionic 
basis employing truncated at $N_F=20$ basis and globally optimized scaling  factors $f$ 
of  oscillator frequency of the basis. In contrast, significantly larger fermionic basis 
reaching $N_F=34$ in superheavy nuclei is required to obtain comparable global 
accuracy of the description of binding energies for  the PC functionals
(see blue curve in Fig.\ \ref{evol-NF}).
Note that the reduction of required accuracy down to $\varepsilon=100$ keV decreases
the required size of basis by only $3-4$ shells (compare red and blue curves in Fig.\ 
\ref{evol-NF}).

\begin{figure}[htb]
    \centering
    \includegraphics*[width=1.00\linewidth]{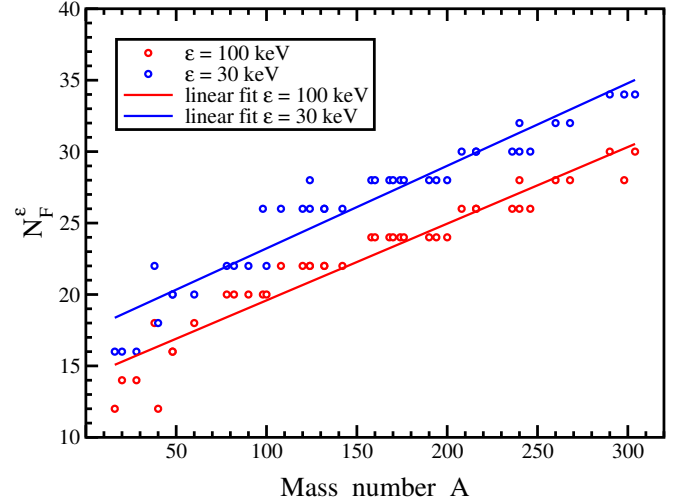}
    \caption{The evolution of the $N_F^{\varepsilon}$ values, providing the numerical
    accuracies of the calculations $\varepsilon=30$ and 100 keV, as a function of
    mass number $A$. Based on the results presented in Table \ref{Table-opt}.   
    \label{evol-NF}
    }
\end{figure}

\section{Halo nuclei in the harmonic oscillator basis}
\label{Halo}

      The finite harmonic oscillator  basis in nuclear many-body 
calculations effectively imposes a  hard-wall boundary conditions in the coordinate 
space, i.e. it is equivalent to a spherical cavity of a radius $L_0$ \cite{FHP.12,BEHPW.16} 
\begin{eqnarray} 
L_0 = \sqrt{2(N_F+3/2)b}.  
\label{radius}
\end{eqnarray} 
in the case of spherical nuclei.  The radius of this cavity is defined by the oscillator 
frequency $\hbar \omega_0$ given by  Eq.\ (\ref{hbar-omega}) and $N_F$ of the 
employed HO basis. Here, 
$b=\sqrt{\hbar/(m\omega_0)}$  is the oscillator length of the basis and $m$ denotes the 
nucleon mass. Note that Eq.\ (\ref{radius})  provides a rough estimate of $L_0$
(see Refs.\ \cite{MEFHP.13,OAD.25}).  
Thus, the radius $L_0$ of spherical cavity for large $N_F$ behaves as 
\begin{eqnarray}
L_0 \sim \sqrt{\frac{N_F}{\sqrt{f}}}
\label{L0-radius} 
\end{eqnarray}
i.e. it increases with decreasing $f$ and increasing $N_F$.

\begin{figure}[htb]
    \centering
    \includegraphics*[width=1.00\linewidth]{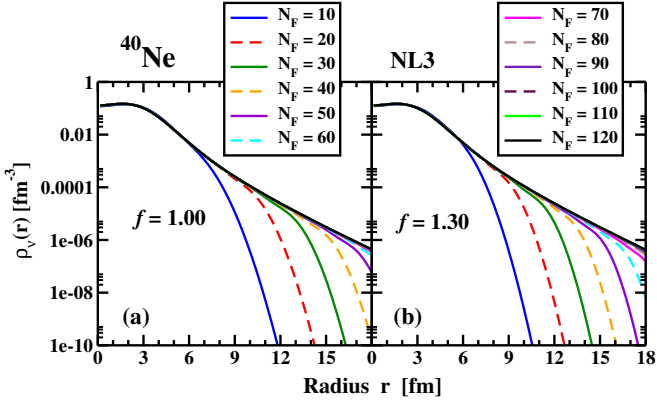}
    \caption{Neutron density in very neutron rich $^{40}$Ne nucleus
    as a function of $N_F$ for two indicated values of scaling factor $f$.  For the
    sake of comparison with coordinate space results,  the
    calculations are carried out with the NL3 functional employed in Ref.\ \cite{PVLR.97}.
    \label{Ne-halo-buildup}
    }
\end{figure}

   Fig.\ \ref{Ne-halo-buildup} shows  sequential buildup of the density in the neutron 
halo and the increase of the radius of spherical cavity
with increasing the $N_F$ value in very neutron-rich $^{40}$Ne nucleus
for two values of scaling factor $f$.  One can see that in both cases the saturation
of the density takes place on approaching $N_F=120$. The $N_F=120$ density is
very well reproduced up to $r=18$ fm in the $N_F=60$ basis for the $f=1.0$
value but requires higher basis of $N_F=80$ in the $f=1.3$ calculations.  This 
is consistent with a general trend defined by Eq.\ (\ref{L0-radius}) which shows
that the radius $L_0$ of spherical cavity increases with decreasing $f$.   Further 
reduction in the basis size required for a reproduction of the $N_F=120$ solutions  
can be achieved by an additional decrease of the $f$  value. However, the analysis 
similar to that presented in Sec. V of Ref.\ \cite{OAD.25} has to be carried out to 
define the lowest value of $f$ below which numerical instabilities in binding energy 
develop.
 
\begin{figure}[htb]
    \centering
    \includegraphics*[width=1.00\linewidth]{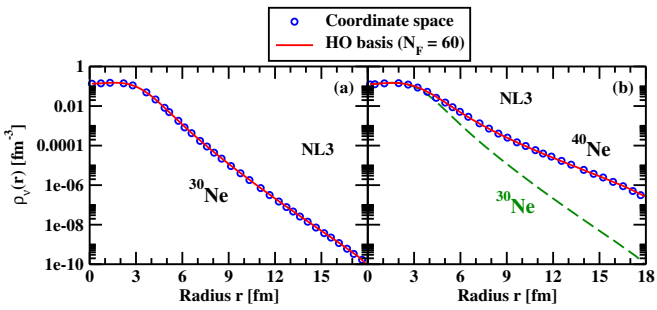}
    \caption{The comparison of  neutron densities  obtained in the HO basis 
    with $N_F=60$ with those generated in coordinate space calculations 
    of Ref.\ \cite{PVLR.97}.   
    \label{Halo-comp}
    }
\end{figure}
 
Neutron densities of the $^{30,40}$Ne nuclei obtained in the HO basis and in coordinate 
space calculations are compared in Fig.\ \ref{Halo-comp}. One can see that the 
$(N_F=60, f=1.0)$ results   very well reproduce  those obtained in 
coordinate space and the mechanism of the formation of neutron halo [i.e. the difference in density between 
$^{30}$Ne and $^{40}$Ne, see Fig.\ \ref{Halo-comp}(b)]. In addition, similar studies have
been carried for heavier $^{72}$Ca nucleus: they also show that the 
densities of neutron halos    obtained  in coordinate space 
calculations of Ref.\ \cite{ZMR.03}  are reproduced with high precision in the HO 
basis with $N_F=120$.

\begin{table*}
	\centering
	\caption{Root mean square deviations $\delta{e_{rms}}$ between the single-particle energies  
	              calculated in the infinite ($N_F=120$) and truncated ($N_F=20$) bases for two values of
	              scaling factor $f$ of oscillator frequency. 
	               The $\delta{e_{rms}}$ values for a given subsystem
	              are presented in the $A/B$ format where $A$ is defined from the energies of all negative energy 
	              single-particle states and $B$ from the energies of only occupied ones.	              
	              The numbers of  negative energy single-particle states  are shown in the "$\#$ of states" columns 
	              in the format $[\#p,\#n]$ where $\#p(\#n)$ is the number of such states in proton (neutron) 
	              subsystem.  
	              \label{Table-single-p}
	                     	            }
\begin{tabular}{cccccccc}\hline \hline
    Scaling     & Nucleus   & \multicolumn{6}{c}{$\delta{e_{rms}}$ [MeV]} \\ \cline{3-8}  
     factor       &   		  & \multicolumn{3}{c}{DD-MEZ} & \multicolumn{3}{c}{PC-Z}  \\ \cline{3-8} 
       $f$         &                    & \# of states &    Proton            & Neutron            & \# of states  &  Proton & Neutron \\ \hline
$f = 1.0$     & $^{48}$Ca   & [10,10]       &  0.007/0.003\,\,  &  0.008/0.004\,\,\,\, &  [10,10]  & 0.007/0.002\,\,  &  0.004/0.002\,\, \\
		    & $^{56}$Ni    & [7,11]        &  0.006/0.006\,\,  &  0.008/0.008\,\,\,\, &  [8,11]  & 0.004/0.003\,\,  &  0.007/0.003\,\, \\
		    & $^{78}$Ni    & [13,13]        &  0.010/0.008\,\,  &  0.035/0.009\,\,\,\, & [13,14]   & 0.013/0.006\,\,  &  0.015/0.009\,\, \\ 
		    & $^{132}$Sn & [16,20]        &  0.011/0.009\,\,  &  0.043/0.013\,\,\,\, &  [16,21]  & 0.020/0.018\,\,  &  0.015/0.010\,\, \\ 
		    & $^{208}$Pb & [19,28]        &  0.018/0.018\,\,  &  0.033/0.023\,\,\,\, &  [19,29]  & 0.018/0.019\,\,  &  0.017/0.016\,\, \\ 
                     &                     &                    &                           &                              &         &                           &                          \\ 
$f^{opt}(A)$  & $^{48}$Ca   & [10,10]        &  0.006/0.002\,\,  &  0.008/0.003\,\,\,\, &  [10,10]  & 0.011/0.001\,\,  &  0.018/0.001\,\, \\ 
		    & $^{56}$Ni     & [7,11]        &  0.004/0.004\,\,  &  0.005/0.004\,\,\,\, &  [8,11]  & 0.006/0.003\,\,  &  0.003/0.003\,\, \\ 
		    & $^{78}$Ni     & [13,13]        &  0.004/0.004\,\,  &  0.058/0.003\,\,\,\, &  [13,14]  & 0.008/0.008\,\,  &  0.091/0.003\,\, \\ 
		    & $^{132}$Sn  & [16,20]        &  0.004/0.004\,\,  &  0.067/0.004\,\,\,\, &  [16,21]  & 0.003/0.003\,\,  &  0.132/0.004\,\, \\ 
		    & $^{208}$Pb  & [19,28]        &  0.007/0.007\,\,  &  0.098/0.006\,\,\,\, &  [19,28]  & 0.008/0.009\,\,  &  0.123/0.009\,\, \\ \hline \hline
	\end{tabular} 
\end{table*}

   This is a consequence of the similarity of the results obtained in very large HO bases 
to those performed in coordinate space. For example, the similarity of the behavior
of positive energy single-particle states in both approaches has been discussed 
in Sec. III.A. of Ref.\ \cite{ALO.26}. To bring the results of both approaches close to each
other it is necessary to increase the radius $L_0$ of effective spherical cavity in the calculations
based on HO basis: this can be done 
by either increasing $N_F$ or by decreasing $f$ or by the combination of both [see
Eq. (\ref{L0-radius})]. 

     Thus, the present study suggests that proton and neutron halos can be 
investigated in theoretical frameworks employing large HO bases. Spherical calculations at 
the mean field level in such bases  are numerically cheap: they require only few minutes 
of the CPU time on a  regular laptop. This is due to moderate  growth of the HO basis  
with  the increase of  $N_F$ (see discussion of Table II in Ref.\ \cite{TOAPT.24}). Thus, 
spherical neutron and proton halos can be safely studied for masses below $A\approx 80$ 
in the computer programs based on the HO basis.  The calculations  of axially deformed
nuclei are significantly more time-consuming (see Ref.\ \cite{TOAPT.24}). However,  relativistic 
calculations (at the RHB level) of such nuclei are feasible in the $N_F\approx 60$, 
$N_F\approx  50$ and $N_F=40$ HO bases in the case of no pairing, of the pairing with only 
diagonal matrix elements (such as monopole one) and of separable pairing of Ref.\ \cite{TMR.09} 
which contains non-diagonal matrix elements (see Ref.\ \cite{NL5Z-DDMEZ-PCZ}). 
Thus, the present results suggest that proton and neutron halos in axially deformed nuclei 
can be studied in the $A\lesssim 40$ nuclei in the framework without pairing. Such structures
can also be studied in the RHB approach with monopole pairing: but in the higher mass end
of this region further reduction of scaling factor $f$ below 1.0 may be required. It is quite likely that 
proton and neutron halos of the axially deformed $A\lesssim 30$ nuclei can be explored in the 
RHB framework with separable pairing but this question requires further investigation.

\section{The consequences of optimization of the harmonic oscillator
              basis for excited states}
\label{excited}

  So far the studies of the optimizations of the HO basis and benchmarking of
their accuracy with respect of infinite basis results have been carried out only for
the ground state properties (see Ref.\ \cite{OAD.25} and Secs. III-V in the
present paper).  However, no such investigations were executed for excited 
single- and multi-particle states and excited collective states in the DFTs
and the results discussed in this section aim at closing this gap in our knowledge.

\subsection{Single-particle states}
\label{excited-sp}

   To understand the accuracy of the description of the single-particle energies
in the HO basis truncated at $N_F$ and whether the optimization of the basis
improves this accuracy it is reasonable to  start from the analysis of a set of 
doubly magic spherical nuclei.  This is because in such nuclei the pairing collapses 
and it is easy to define infinite basis solutions. Root mean square deviations 
$\delta{e_{rms}}$ between the single-particle energies calculated in the infinite and 
truncated bases  are presented for such a set in Table \ref{Table-single-p}.
  Note that  the calculations have been performed both with $f=1.0$ and optimized
$f_{opt}(A)$ values of scaling factor of oscillator frequency. The latter is 
defined by Eq.\ (\ref{f-mass-dep}) for the PC-Z functional and by Eq. (9) and Table 
II of Ref.\ \cite{OAD.25} for the DD-MEZ one.

\begin{figure}[htb]
    \centering
    \includegraphics*[width=8.6cm]{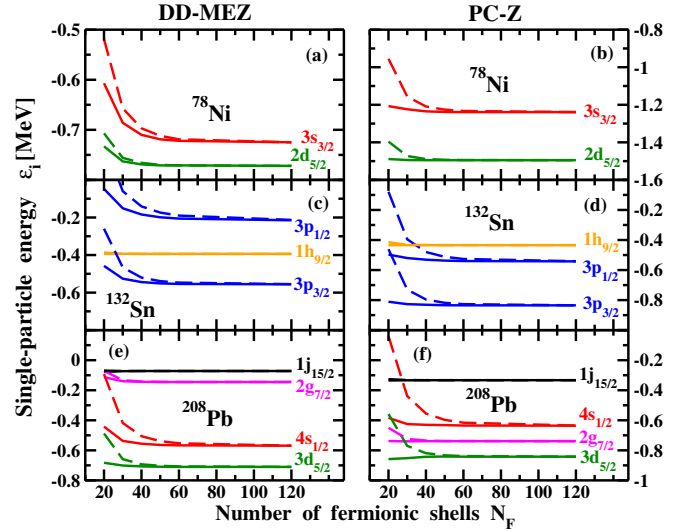}
    \caption{The dependence of the energies of weakly bound single-particle states
                  on $N_F$.  Solid and dashed lines are used for the results obtained with
                  $f=1.0$ and $f_{opt}(A)$, respectively. Red, blue, green, magenta, orange
                  and black colors are used for the single-particle states with orbital angular momentum
                  $l=0$, 1,  2, 4, 5 and 7, respectively.     
                  \label{sp-weakly}
    }
\end{figure}

     One can see in Table \ref{Table-single-p} that for a given functional  the $\delta{e_{rms}}$ 
values obtained for the full set of bound neutron single-particle states and the subset restricted 
to  only occupied neutron states are different. This difference  is especially pronounced in the 
calculations with $f^{opt}(A)$ for $^{78}$Ni, $^{132}$Sn and $^{208}$Pb.
The origin of this feature lies in the fact that the accuracy of the description of the energies
of  weakly bound neutron states with low value of orbital angular momentum of $l=0$, 1 and 2
(see Fig.\ \ref{sp-weakly}) substantially depends on two factors. First, these neutron states are 
located in the energy range which is characterized by a larger  radius of the neutron potential as 
compared with  that for more bound single-particle states (see, for example, Figs. 5(c) and (d)  
and  Figs. 8(c)  and (d) in Ref.\ \cite{PA.22}). As a result, their accurate description requires 
a larger  radius of $L_0$ of spherical cavity discussed in Sec.\ \ref{Halo} which can be achieved
either by the increase of the size of the basis (i.e. $N_F$) or by reducing the scaling factor $f$.
Indeed the basis with $f=1.0$ provides much more accurate description of the energies of neutron 
states with low $l$ at $N_F=20$ and faster convergence to the infinite basis solutions as compared 
with the results obtained with $f_{opt}(A)$ (see Fig.\ \ref{sp-weakly}). Second, the sensitivity of the 
energies of weakly bound neutron states to $N_F$ and $f$ decreases with increasing orbital 
angular momentum. This is clearly seen in the case of the $l=5$ ($1h_{9/2}$) and $l=7 (1j_{15/2}$) 
states: their energies almost do not depend  on these parameters [see Fig.\ \ref{sp-weakly}(c), (d), 
(e) and (f)]. These features reveal the asymptotic properties  of the wave functions of weakly-bound 
low-$l$ neutron orbitals at large radius which are substantially different from those of medium and 
high $l$ orbitals. 
 
    In contrast, these features are not active in the proton subsystem: for a given functional  
the  $\delta{e_{rms}}$ values obtained for the full set of bound proton single-particle states and the 
subset restricted to only occupied proton states are almost the same (see Table \ref{Table-single-p}).
This is due to the fact that the properties of proton potential are substantially different from those
of the neutron one due to the presence of Coulomb potential (for example, compare proton and 
neutron potentials in Figs. 5 and 7 of Ref.\ \cite{PA.22}). 

    If to exclude low-$l$ weakly bound neutron states from the consideration\footnote{
Such states will not affect the properties of almost all nuclei 
located reasonably far away from the neutron drip line since they lie far 
away in energy from the Fermi level (see, for example, neutron 
single-particle diagrams shown in Figs. 2, 3, 4 and 5 of  Ref.\ \cite{LA.11}).
As a result, they are either not occupied or occupied with marginal 
probability in the calculations with pairing.} 
      then in the calculations
with the DD-MEZ functional  the $\delta{e_{rms}}$  values for negative energy states decrease from 
0.058 MeV in $^{78}$Ni,  0.067 MeV in $^{132}$Sn and 0.098 MeV in $^{208}$Pb obtained in the calculations 
with $f_{opt}(A)$ (see Table \ref{Table-single-p}) down to 0.003, 0.012  and 0.016 MeV. These values 
are either the same ($^{78}$Ni) or are reasonably close ($^{132}$Sn and $^{208}$Pb)  to those 
obtained for only occupied single-particle states (see Table \ref{Table-single-p}). A similar situation
exists  in the calculations with PC-Z. 

  Based on the results presented for only occupied negative energy states (column B in the A/B format) 
in Table \ref{Table-single-p} one can conclude that the optimization of the basis improves the accuracy 
of the description of  the energies of the single-particle states by a factor of approximately  two 
[compare the results obtained with $f=1.0$ and $f_{opt}(A)$].   In all cases, it is better than 10 
keV in the calculations with $f_{opt}(A)$. The same accuracy is expected for the ground and 
excited single-particle states in odd-$A$ nuclei.  The accuracy of the description of the energies of 
two- and/or multi-particle configurations can be easily defined by using the standard rules
of the propagation of errors and assuming that each involved state is described with an 
accuracy $\delta{e_{rms}}$ provided in Table \ref{Table-single-p}. 

The deformation of the nucleus leads to the mixing of the wave functions of the single-particle states in the 
nuclei. Thus, for moderate deformation of the nuclei (up to $\beta_2 \approx 0.3-04$) it is reasonable to 
expect that the rms errors in the description of the energies of the set of deformed single-particle states 
emerging from the  set of spherical subshells studied in Table \ref{Table-single-p}  will be comparable to the 
rms errors presented in the quoted table. However, these errors may increase for the single-particle states
at superdeformation since intruder orbitals from higher lying shells become either occupied or located
close in energy to the Fermi level.  This substantially affects the mixing of the wave functions
of  the single-particle states (see, for example, Ref.\ \cite{AA.18}).

 Note that the considerations in the previous two paragraphs are strictly valid only for the nuclei 
which are located reasonably far away from the neutron drip line and the configurations of which 
do not involve the occupation of low-$l$ weakly bound neutron single-particle states. More than
95\%  of experimentally known nuclei belong to this category and these nuclei are used in the global 
fits of EDFs.  Similar to Ref.\ \cite{OAD.25} the present paper is focused on the optimization of the 
HO basis for such nuclei which are most frequently studied experimentally and theoretically. However, 
the results presented in Sec.\ \ref{Halo} and 
in Fig.\ \ref{sp-weakly} strongly suggests  that the optimization of the HO basis can be quite 
different in very neutron-rich nuclei and, in  particular, in the neutron halo ones. For example, 
lower values of scaling factor $f$ as compared with $f_{opt}(A)$ can be more favored in such 
nuclei. Moreover, the  proper optimization of the HO basis in such nuclei is further complicated by 
the fact that the differences in neutron and proton density distributions increase with approaching 
the neutron drip line. This may require  the introduction of different oscillator frequencies  for proton 
and neutron subsystems which will make the optimization of the basis more complex.

\subsection{Fission barriers and fission isomers}
\label{excited-fis}

  The fission process sensitively depends on numerical accuracy of the predictions 
of the height of fission barrier. For example, the modification of fission barrier
by 1 MeV changes the calculated spontaneous fission half-lives by many orders
of magnitude (see Refs.\ \cite{BKRRSW.15,SR.16}).  Typically, the accuracy of the calculation of fission 
barrier in a given truncation of basis is estimated by some increase of the basis 
(for example,  by increasing the size of the basis by $\Delta N_F=2$). However, 
this is a relatively crude approach since true numerical uncertainties cannot 
be evaluated in this way. To do that they have to be estimated with respect 
of either the results obtained in infinite fermionic basis or those which
reasonably well approximate such a basis.  However, no such evaluation is
available in the literature. Another question is how numerical uncertainties
in the calculations of fission barriers depend on the selection of scaling 
factor $f$ of oscillator frequency $\hbar \omega_0$ of the basis. To our
knowledge this question has also not been addressed in the literature.

\begin{figure*}[htb]
    \centering
    \includegraphics*[width=0.8\linewidth]{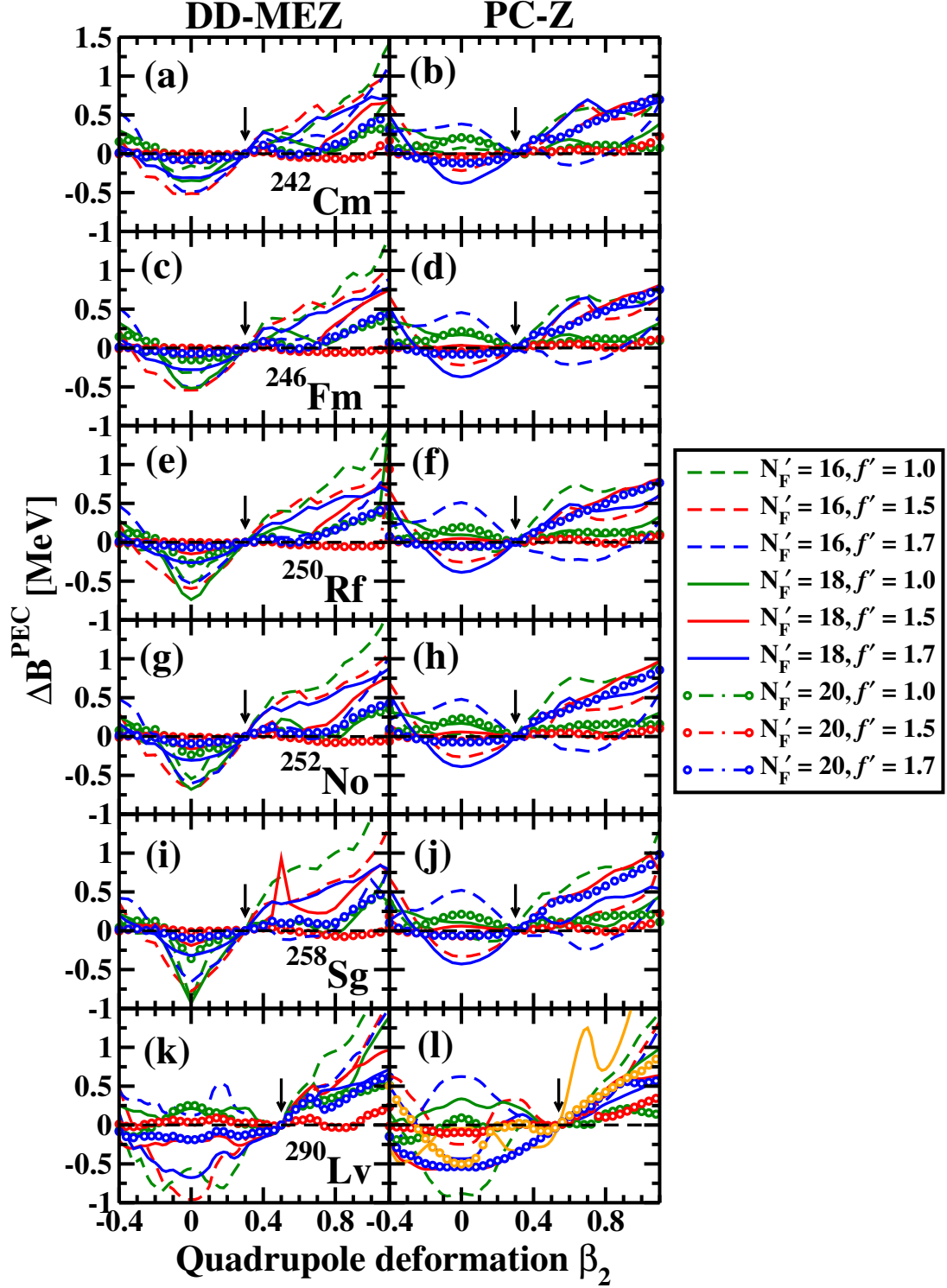}
    \caption{
     The evolution of the $\Delta B^{PEC}$ quantity as a function of quadrupole 
     deformation $\beta_2$ for selected set of the $(N_F',f')$ basis truncations
     in indicated nuclei.  Black dashed line shows   $\Delta B^{PEC}=0.0$.
     The vertical black arrow shows the position of the global minimum.  The 
     $(N_F'=18, f'=2.0)$ and $(N_F'=20, f'=2.0)$ results are shown 
     in panel (l) by orange solid line and orange open circles, respectively.        
    \label{fission-barrier-summary}
    }
\end{figure*}

   To address this gap in our knowledge we calculate the $\Delta B^{PEC}$ quantity
defined as
\begin{eqnarray}
\Delta B^{PEC} & = & \left[B_{N_F=34,f=1.5}(\beta_2)-B_{N_F=34,f=1.5}(\beta_2^{min})\right] -  \nonumber \\              
 &&-\left[B_{N_F',f'}(\beta_2)-B_{N_F',f'}(\beta_2^{min})\right]
\end{eqnarray} 
where $\beta_2^{min}$ is quadrupole deformation of the global minimum. It 
compares potential energy curves (PEC) calculated in the $(N_F', f')$ basis 
with those obtained in the $(N_F=34, f=1.5)$ one. The PECs obtained in the 
latter basis serve as a benchmark closely approximating the results obtained in
infinite fermionic basis.  Thus, $\Delta B^{PEC}$ represents a measure of
the accuracy of the reproduction of the PECs calculated in the $(N_F', f')$ basis
as compared with the infinite one.

   The selection of the $f=1.5$ value in the $N_F=34$ calculations is guided by two  factors.
First, this is an optimal value of the scaling factor $f$ for the description of  the ground state 
properties with the DD-MEZ functional (see Ref.\ \cite{OAD.25}). It also provides a reasonable
convergence for the PECs  in constrained calculations. Second, this value is somewhat lower 
as compared with the optimal value of $f=1.7-1.8$ recommended 
for deformed nuclei in the PC-Z functional (see Table \ref{Table-opt}) but the numerical 
calculations of the PECs are faster and more stable for $f=1.5$ than for recommended
value of $f$.  The $N_F=34, f=1.5$ results are reasonably close to those extrapolated 
to the infinite basis: thus, for simplicity of discussion, we label them further as EIB 
(extrapolated to the infinite basis). For example, the comparison of the results obtained
with $N_F=34$ and $N_F'=38$ at some deformation points 
of the PECs   in selected set of nuclei
shows that $|\Delta B^{PEC}|$ values are below 3 keV in most of the 
cases\footnote{The comparison of full PECs in these two truncations 
of basis is not possible since the convergence in the region of the
fission barrier saddle requires substantially more iterations than that for  deformation 
points in the vicinity of normal and superdeformed minima.  As a result, 
the time of numerical calculations for a number of deformation points in the $N_F=38$ 
basis exceeds the limit existing on available computers.}.
Moreover, the difference in the $\Delta B^{PEC}$ values obtained in 
the  $N_F=34$ calculations with $f=1.0$ and $f=1.5$ is on average smaller than 10 keV 
with the maximum value being less than 20 keV.

    Fig.\ \ref{fission-barrier-summary}  compares PECs calculated in different $(N_F',f')$
truncation schemes for a selected set of actinides and superheavy nuclei in the 
deformation range from $\beta_2=-0.4$ up to $\beta_2=1.1$. The upper limit of $\beta_2$
is defined by the fact that for larger deformations numerical calculations in the $N_F=34$
basis exceed time limit on available computers. Thus, above mentioned range covers the
deformations typical for shape coexistence, inner fission barrier and fission isomers located
in the superdeformed  minimum (see Ref.\ \cite{TAA.20}).

   A number of important conclusions can be drawn from the analysis of Fig.\ 
\ref{fission-barrier-summary}.  First, the $N_F=16$ basis generates quite large 
errors typically exceeding 1.0 MeV in the DD-MEZ CEDF and 0.5 MeV in the PC-Z 
one at the upper limit of deformation range studied. These errors in the description
of EIB PECs are also quite substantial (reaching $0.5-1.0$ MeV) even at low $\beta_2$ deformations 
located between $-0.4$ and 0.4.  There is also a sizable dependence of the 
$\Delta B^{PEC}$ curves on the value of scaling factor $f$.  The use of the $N_F=18$ 
basis somewhat improves the situation and on average the numerical errors in the 
description of PECs reduce as compared with the $N_F=16$  case. However, above 
mentioned features still persist.

  In general, the transition to the $N_F=20$ basis leads to a further reduction of numerical 
errors in the description of EIB potential energy curves and this effect in especially pronounced in the 
case of the DD-MEZ functional (see Fig.\ \ref{fission-barrier-summary}). However, even 
then there is a substantial sensitivity of the accuracy of the reproduction of the EIB 
solutions to the scaling factor $f$. It is only the combination of $N_F'=20$ and $f'=1.5$ 
which reproduces the EIB solutions with an accuracy that is better than 100 keV. The only 
exception  is the case of high deformations in $^{290}$Lv where the errors approach 
$\approx 250$ keV at $\beta_2=1.1$ [see Figs.\ \ref{fission-barrier-summary}(k) and (l)]. 
The errors in the reproduction of EIB PECs increase when other values of scaling factor $f$
such as $f'=1.0$, $f'=1.7$ and $f'=2.0$ are used (see Fig.\ \ref{fission-barrier-summary}).

   The present analysis clearly indicates that the optimization of the HO basis can substantially 
improve the accuracy of the description of potential energy curves and thus the energies of the 
saddles of the fission barriers and the excitation energies of the fission isomers.  However, even 
then the size of the basis has to be substantial since it is only at $N_F=20$ that it is possible to 
find a value of scaling factor $f$ of oscillator frequency $\hbar\omega_0$ (i.e. $f=1.5$)  which accurately (within 100 
keV) and consistently (with exception of high deformation  region of $^{290}$Lv) reproduces EIB 
potential energy curves of the actinides and  superheavy nuclei. This value of $f$ is consistent 
with the one obtained for meson exchange functionals in the  optimization  of the HO basis with 
respect of binding energies of the ground states (see Ref.\ \cite{OAD.25}) but it is somewhat smaller 
than the one ($f=1.7-1.8)$ obtained for point coupling functionals in the present paper. 

\section{Conclusions}
\label{Concl} 

  The main goal of the present study is further development of covariant density 
functional theory toward a more accurate description of physical observables of
interest within a moderately sized fermionic basis. The present study is focused on the 
improvement of the description of both the ground state (binding energies, charge 
radii and halo nuclei) and excited state (fission barriers, fission isomers and single-particle
energies) properties.  This is achieved by an optimization of the truncated $(N_F,f)$ basis 
with respect of infinite basis solutions and by a better understanding of the convergence 
properties of the former.  A special attention is paid to the point coupling covariant 
energy density functionals.
 
 The main results can be summarized as follows.

\begin{itemize}

\item
  The convergence of binding energies as a function of $N_F$ to $B(N_F=\infty)$
for point coupling functionals improves drastically in the optimized $f_{opt}(A)$ basis 
as compared with the $f=1.0$ one.  This allows to resolve the issue with the definition
of extrapolated infinite basis solutions for actinides and superheavy nuclei which 
existed for this class of the CEDFs  in earlier studies of Refs.\ 
\cite{TOAPT.24,NL5Z-DDMEZ-PCZ}.   The recommended $f_{opt}(A)$ values of scaling 
factor $f$ of oscillator frequency $\hbar\omega_0$ and recommended sizes of the basis 
$N_F^{\varepsilon}(A)$ for controllable errors $\varepsilon=0.1$ and 0.03 MeV are provided 
for experimentally known part of nuclear chart (see Fig.\ \ref{map-of-nuclei}). The size of 
the basis required to reach a given level of accuracy in the description of binding energies 
for the point coupling functionals is substantially larger than the one for meson exchange 
ones.
 
\item 
  Non-regular changes in the densities with  increasing  $N_F$ cause self-consistent 
feedback to binding energies which leads to non-regular fluctuations of these 
energies around some smooth trend with increasing $N_F$, i.e. to odd-even staggering 
in convergence curves of binding energies. This effect is expected to be present also in
non-relativistic density functional theories but absent in phenomenological potentials.
These non-regular fluctuations in binding energies prevent a reliable extrapolation at 
moderate sizes of  truncated basis to the infinite one by the extrapolation methods such 
as Shanks transformation.

\item
    The calculations in very large fermionic HO bases reproduce neutron
halo densities obtained in the coordinate space calculations. This 
illustrates the applicability of the approaches based on the expansion 
of  wave functions into the HO basis states to the study of the physics 
of halos in nuclei. It is estimated that the RHB computer codes adapted 
to very large HO bases in Ref.\ \cite{NL5Z-DDMEZ-PCZ} allow to study 
the halo structures  in spherical and axially deformed nuclei up to 
$A\approx 80$ and 40, respectively.

\item
   The optimization of the HO basis can  substantially improve the accuracy of 
the description of potential energy curves. However, this requires a substantial 
size of the HO basis such as $N_F=20$. For example, the 100 keV accuracy in 
the description of potential energy curves extrapolated to the infinite basis is 
reached in the $(N_F=20, f=1.5)$ basis. This also defines the accuracy of the
description of the energies of the  saddles  of the fission barriers and the excitation 
energies of  the fission  isomers.

\item
  The accuracy of the description of  the energies of the single-particle states 
is improved by a factor of approximately  two when optimized HO basis is used. 
It is better than 10 keV for the single-particle states  of doubly magic spherical nuclei 
investigated in the present paper.  The same accuracy  is expected for the ground 
and  excited single-particle states in odd-$A$ spherical and deformed nuclei.  The 
only exception  from this rule are weakly bound neutron single-particle states with 
low orbital angular  momentum $l$ which favor the description in the basis with 
$f<f^{opt}(A)$.
\\
  
\end{itemize}

\section{ACKNOWLEDGMENTS}

 This material is based upon work supported by the U.S. Department of Energy,  
Office of Science, Office of Nuclear Physics under Award No. DE-SC0013037.

\bibliography{references-48-optimized-basis.bib}

\begin{thebibliography}{36}%
\makeatletter
\providecommand \@ifxundefined [1]{%
 \@ifx{#1\undefined}
}%
\providecommand \@ifnum [1]{%
 \ifnum #1\expandafter \@firstoftwo
 \else \expandafter \@secondoftwo
 \fi
}%
\providecommand \@ifx [1]{%
 \ifx #1\expandafter \@firstoftwo
 \else \expandafter \@secondoftwo
 \fi
}%
\providecommand \natexlab [1]{#1}%
\providecommand \enquote  [1]{``#1''}%
\providecommand \bibnamefont  [1]{#1}%
\providecommand \bibfnamefont [1]{#1}%
\providecommand \citenamefont [1]{#1}%
\providecommand \href@noop [0]{\@secondoftwo}%
\providecommand \href [0]{\begingroup \@sanitize@url \@href}%
\providecommand \@href[1]{\@@startlink{#1}\@@href}%
\providecommand \@@href[1]{\endgroup#1\@@endlink}%
\providecommand \@sanitize@url [0]{\catcode `\\12\catcode `\$12\catcode
  `\&12\catcode `\#12\catcode `\^12\catcode `\_12\catcode `\%12\relax}%
\providecommand \@@startlink[1]{}%
\providecommand \@@endlink[0]{}%
\providecommand \url  [0]{\begingroup\@sanitize@url \@url }%
\providecommand \@url [1]{\endgroup\@href {#1}{\urlprefix }}%
\providecommand \urlprefix  [0]{URL }%
\providecommand \Eprint [0]{\href }%
\providecommand \doibase [0]{https://doi.org/}%
\providecommand \selectlanguage [0]{\@gobble}%
\providecommand \bibinfo  [0]{\@secondoftwo}%
\providecommand \bibfield  [0]{\@secondoftwo}%
\providecommand \translation [1]{[#1]}%
\providecommand \BibitemOpen [0]{}%
\providecommand \bibitemStop [0]{}%
\providecommand \bibitemNoStop [0]{.\EOS\space}%
\providecommand \EOS [0]{\spacefactor3000\relax}%
\providecommand \BibitemShut  [1]{\csname bibitem#1\endcsname}%
\let\auto@bib@innerbib\@empty
\bibitem [{\citenamefont {Osei}\ \emph
  {et~al.}(2025{\natexlab{a}})\citenamefont {Osei}, \citenamefont {Afanasjev},\
  and\ \citenamefont {Dalbah}}]{OAD.25}%
  \BibitemOpen
  \bibfield  {author} {\bibinfo {author} {\bibfnamefont {B.}~\bibnamefont
  {Osei}}, \bibinfo {author} {\bibfnamefont {A.~V.}\ \bibnamefont
  {Afanasjev}},\ and\ \bibinfo {author} {\bibfnamefont {A.}~\bibnamefont
  {Dalbah}},\ }\bibfield  {title} {\bibinfo {title} {Global optimization of
  harmonic oscillator basis in covariant density functional theory},\
  }\href@noop {} {\bibfield  {journal} {\bibinfo  {journal} {Phys. Rev. C}\
  }\textbf {\bibinfo {volume} {112}},\ \bibinfo {pages} {054321} (\bibinfo
  {year} {2025}{\natexlab{a}})}\BibitemShut {NoStop}%
\bibitem [{\citenamefont {Coon}\ \emph {et~al.}(2012)\citenamefont {Coon},
  \citenamefont {Avetian}, \citenamefont {Kruse}, \citenamefont {van Kolck},
  \citenamefont {Maris},\ and\ \citenamefont {Vary}}]{CAKKMV.12}%
  \BibitemOpen
  \bibfield  {author} {\bibinfo {author} {\bibfnamefont {S.~A.}\ \bibnamefont
  {Coon}}, \bibinfo {author} {\bibfnamefont {M.~I.}\ \bibnamefont {Avetian}},
  \bibinfo {author} {\bibfnamefont {M.~K.~G.}\ \bibnamefont {Kruse}}, \bibinfo
  {author} {\bibfnamefont {U.}~\bibnamefont {van Kolck}}, \bibinfo {author}
  {\bibfnamefont {P.}~\bibnamefont {Maris}},\ and\ \bibinfo {author}
  {\bibfnamefont {J.~P.}\ \bibnamefont {Vary}},\ }\bibfield  {title} {\bibinfo
  {title} {Convergence properties of {\it ab initio} calculations of light
  nuclei in a harmonic oscillator basis},\ }\href@noop {} {\bibfield  {journal}
  {\bibinfo  {journal} {Phys. Rev. C}\ }\textbf {\bibinfo {volume} {86}},\
  \bibinfo {pages} {054002} (\bibinfo {year} {2012})}\BibitemShut {NoStop}%
\bibitem [{\citenamefont {Gambhir}\ \emph {et~al.}(1990)\citenamefont
  {Gambhir}, \citenamefont {Ring},\ and\ \citenamefont {Thimet}}]{GRT.90}%
  \BibitemOpen
  \bibfield  {author} {\bibinfo {author} {\bibfnamefont {Y.~K.}\ \bibnamefont
  {Gambhir}}, \bibinfo {author} {\bibfnamefont {P.}~\bibnamefont {Ring}},\ and\
  \bibinfo {author} {\bibfnamefont {A.}~\bibnamefont {Thimet}},\ }\bibfield
  {title} {\bibinfo {title} {Relativistic mean field theory for finite
  nuclei},\ }\href@noop {} {\bibfield  {journal} {\bibinfo  {journal} {Ann.\
  Phys. (N.Y.)}\ }\textbf {\bibinfo {volume} {198}},\ \bibinfo {pages} {132}
  (\bibinfo {year} {1990})}\BibitemShut {NoStop}%
\bibitem [{\citenamefont {Nilsson}\ and\ \citenamefont
  {Ragnarsson}()}]{NilRag-book}%
  \BibitemOpen
  \bibfield  {author} {\bibinfo {author} {\bibfnamefont {S.~G.}\ \bibnamefont
  {Nilsson}}\ and\ \bibinfo {author} {\bibfnamefont {I.}~\bibnamefont
  {Ragnarsson}},\ }\bibfield  {title} {\bibinfo {title} {Shapes and shells in
  nuclear structure},\ }\href@noop {} {\bibinfo  {journal} {{\it Shapes and
  shells in nuclear structure}, (Cambridge University Press, 1995)}\
  }\BibitemShut {NoStop}%
\bibitem [{\citenamefont {Kvaal}(2009)}]{K.2009}%
  \BibitemOpen
\bibfield  {journal} {  }\bibfield  {author} {\bibinfo {author} {\bibfnamefont
  {S.}~\bibnamefont {Kvaal}},\ }\bibfield  {title} {\bibinfo {title} {Harmonic
  oscillator eigernfunction expansion, quantum dots, and effective
  interactions},\ }\href@noop {} {\bibfield  {journal} {\bibinfo  {journal}
  {Phys. Rev. B}\ }\textbf {\bibinfo {volume} {88}},\ \bibinfo {pages} {045321}
  (\bibinfo {year} {2009})}\BibitemShut {NoStop}%
\bibitem [{\citenamefont {Taninah}\ \emph {et~al.}(2024)\citenamefont
  {Taninah}, \citenamefont {Osei}, \citenamefont {Afanasjev}, \citenamefont
  {Perera},\ and\ \citenamefont {Teeti}}]{TOAPT.24}%
  \BibitemOpen
  \bibfield  {author} {\bibinfo {author} {\bibfnamefont {A.}~\bibnamefont
  {Taninah}}, \bibinfo {author} {\bibfnamefont {B.}~\bibnamefont {Osei}},
  \bibinfo {author} {\bibfnamefont {A.~V.}\ \bibnamefont {Afanasjev}}, \bibinfo
  {author} {\bibfnamefont {U.~C.}\ \bibnamefont {Perera}},\ and\ \bibinfo
  {author} {\bibfnamefont {S.}~\bibnamefont {Teeti}},\ }\bibfield  {title}
  {\bibinfo {title} {Toward accurate nuclear mass tables in covariant density
  functional theory},\ }\href@noop {} {\bibfield  {journal} {\bibinfo
  {journal} {Phys. Rev. C}\ }\textbf {\bibinfo {volume} {109}},\ \bibinfo
  {pages} {024321} (\bibinfo {year} {2024})}\BibitemShut {NoStop}%
\bibitem [{\citenamefont {Furnstahl}\ \emph {et~al.}(2012)\citenamefont
  {Furnstahl}, \citenamefont {Hagen},\ and\ \citenamefont
  {Papenbrock}}]{FHP.12}%
  \BibitemOpen
  \bibfield  {author} {\bibinfo {author} {\bibfnamefont {R.~N.}\ \bibnamefont
  {Furnstahl}}, \bibinfo {author} {\bibfnamefont {G.}~\bibnamefont {Hagen}},\
  and\ \bibinfo {author} {\bibfnamefont {T.}~\bibnamefont {Papenbrock}},\
  }\bibfield  {title} {\bibinfo {title} {Corrections to nuclear energies and
  radii in finite oscillator spaces},\ }\href@noop {} {\bibfield  {journal}
  {\bibinfo  {journal} {Phys. Rev. C}\ }\textbf {\bibinfo {volume} {86}},\
  \bibinfo {pages} {031301(R)} (\bibinfo {year} {2012})}\BibitemShut {NoStop}%
\bibitem [{\citenamefont {More}\ \emph {et~al.}(2013)\citenamefont {More},
  \citenamefont {Ekstr{\"o}m}, \citenamefont {Furnstahl}, \citenamefont
  {Hagen},\ and\ \citenamefont {Papenbrock}}]{MEFHP.13}%
  \BibitemOpen
  \bibfield  {author} {\bibinfo {author} {\bibfnamefont {S.}~\bibnamefont
  {More}}, \bibinfo {author} {\bibfnamefont {A.}~\bibnamefont {Ekstr{\"o}m}},
  \bibinfo {author} {\bibfnamefont {R.~J.}\ \bibnamefont {Furnstahl}}, \bibinfo
  {author} {\bibfnamefont {G.}~\bibnamefont {Hagen}},\ and\ \bibinfo {author}
  {\bibfnamefont {T.}~\bibnamefont {Papenbrock}},\ }\bibfield  {title}
  {\bibinfo {title} {Universal properties of infrared oscillator basis
  extrapolations},\ }\href@noop {} {\bibfield  {journal} {\bibinfo  {journal}
  {Phys. Rev. C}\ }\textbf {\bibinfo {volume} {87}},\ \bibinfo {pages} {044326}
  (\bibinfo {year} {2013})}\BibitemShut {NoStop}%
\bibitem [{\citenamefont {Binder}\ \emph {et~al.}(2016)\citenamefont {Binder},
  \citenamefont {Ekstr{\"o}m}, \citenamefont {Hagen}, \citenamefont
  {Papenbrock},\ and\ \citenamefont {Wendt}}]{BEHPW.16}%
  \BibitemOpen
  \bibfield  {author} {\bibinfo {author} {\bibfnamefont {S.}~\bibnamefont
  {Binder}}, \bibinfo {author} {\bibfnamefont {A.}~\bibnamefont {Ekstr{\"o}m}},
  \bibinfo {author} {\bibfnamefont {G.}~\bibnamefont {Hagen}}, \bibinfo
  {author} {\bibfnamefont {T.}~\bibnamefont {Papenbrock}},\ and\ \bibinfo
  {author} {\bibfnamefont {K.~A.}\ \bibnamefont {Wendt}},\ }\bibfield  {title}
  {\bibinfo {title} {Effective filed theory in the harmonic oscillator basis},\
  }\href@noop {} {\bibfield  {journal} {\bibinfo  {journal} {Phys. Rev. C}\
  }\textbf {\bibinfo {volume} {93}},\ \bibinfo {pages} {044332} (\bibinfo
  {year} {2016})}\BibitemShut {NoStop}%
\bibitem [{\citenamefont {Coon}\ and\ \citenamefont {Kruse}(2016)}]{CK.16}%
  \BibitemOpen
  \bibfield  {author} {\bibinfo {author} {\bibfnamefont {S.~A.}\ \bibnamefont
  {Coon}}\ and\ \bibinfo {author} {\bibfnamefont {M.~K.~G.}\ \bibnamefont
  {Kruse}},\ }\bibfield  {title} {\bibinfo {title} {Properties of infrared
  extrapolations in a harmonic oscillator basis},\ }\href@noop {} {\bibfield
  {journal} {\bibinfo  {journal} {Int. J. Mod. Phys. E}\ }\textbf {\bibinfo
  {volume} {25}},\ \bibinfo {pages} {1641011} (\bibinfo {year}
  {2016})}\BibitemShut {NoStop}%
\bibitem [{\citenamefont {Mazur}\ \emph {et~al.}(2025)\citenamefont {Mazur},
  \citenamefont {Sharypov},\ and\ \citenamefont {Shirokov}}]{MSS.25}%
  \BibitemOpen
  \bibfield  {author} {\bibinfo {author} {\bibfnamefont {A.}~\bibnamefont
  {Mazur}}, \bibinfo {author} {\bibfnamefont {R.}~\bibnamefont {Sharypov}},\
  and\ \bibinfo {author} {\bibfnamefont {A.}~\bibnamefont {Shirokov}},\
  }\bibfield  {title} {\bibinfo {title} {Extrapolation to infinite model space
  of no-core shell model calculations using machine learning},\ }\href@noop {}
  {\bibfield  {journal} {\bibinfo  {journal} {nuclear theory archieve
  arXiv:2511.05061v2}\ } (\bibinfo {year} {2025})}\BibitemShut {NoStop}%
\bibitem [{\citenamefont {Kn{\"o}ll}\ \emph {et~al.}(2025)\citenamefont
  {Kn{\"o}ll}, \citenamefont {Lockner}, \citenamefont {Maris}, \citenamefont
  {McCarty}, \citenamefont {Roth}, \citenamefont {Vary},\ and\ \citenamefont
  {Wolfgruber}}]{KLMMRVW.25}%
  \BibitemOpen
  \bibfield  {author} {\bibinfo {author} {\bibfnamefont {M.}~\bibnamefont
  {Kn{\"o}ll}}, \bibinfo {author} {\bibfnamefont {M.}~\bibnamefont {Lockner}},
  \bibinfo {author} {\bibfnamefont {P.}~\bibnamefont {Maris}}, \bibinfo
  {author} {\bibfnamefont {R.~J.}\ \bibnamefont {McCarty}}, \bibinfo {author}
  {\bibfnamefont {R.}~\bibnamefont {Roth}}, \bibinfo {author} {\bibfnamefont
  {J.~P.}\ \bibnamefont {Vary}},\ and\ \bibinfo {author} {\bibfnamefont
  {T.}~\bibnamefont {Wolfgruber}},\ }\bibfield  {title} {\bibinfo {title}
  {Benchmarking ann extrapolations of the ground-state energies and radii of li
  isotopes},\ }\href@noop {} {\bibfield  {journal} {\bibinfo  {journal}
  {nuclear theory archieve arXiv:2501.18252v2}\ } (\bibinfo {year}
  {2025})}\BibitemShut {NoStop}%
\bibitem [{\citenamefont {Dobaczewski}\ and\ \citenamefont
  {Dudek}(1997)}]{DD.97}%
  \BibitemOpen
  \bibfield  {author} {\bibinfo {author} {\bibfnamefont {J.}~\bibnamefont
  {Dobaczewski}}\ and\ \bibinfo {author} {\bibfnamefont {J.}~\bibnamefont
  {Dudek}},\ }\bibfield  {title} {\bibinfo {title} {Solution of the
  skyrme-hartree-fock equations in the cartesian deformed harmonic oscillator
  basis. i. the method},\ }\href@noop {} {\bibfield  {journal} {\bibinfo
  {journal} {Comp. Phys. Comm.}\ }\textbf {\bibinfo {volume} {102}},\ \bibinfo
  {pages} {166} (\bibinfo {year} {1997})}\BibitemShut {NoStop}%
\bibitem [{\citenamefont {Pillet}\ and\ \citenamefont {Hilaire}(2017)}]{PH.17}%
  \BibitemOpen
  \bibfield  {author} {\bibinfo {author} {\bibfnamefont {N.}~\bibnamefont
  {Pillet}}\ and\ \bibinfo {author} {\bibfnamefont {S.}~\bibnamefont
  {Hilaire}},\ }\bibfield  {title} {\bibinfo {title} {Towards an extended gogny
  force},\ }\href {https://doi.org/10.1140/epja/i2017-12369-3} {\bibfield
  {journal} {\bibinfo  {journal} {Eur. Phys. J. A}\ }\textbf {\bibinfo {volume}
  {53}},\ \bibinfo {pages} {193} (\bibinfo {year} {2017})}\BibitemShut
  {NoStop}%
\bibitem [{\citenamefont {Afanasjev}\ \emph {et~al.}(1996)\citenamefont
  {Afanasjev}, \citenamefont {K\"onig},\ and\ \citenamefont {Ring}}]{AKR.96}%
  \BibitemOpen
  \bibfield  {author} {\bibinfo {author} {\bibfnamefont {A.~V.}\ \bibnamefont
  {Afanasjev}}, \bibinfo {author} {\bibfnamefont {J.}~\bibnamefont {K\"onig}},\
  and\ \bibinfo {author} {\bibfnamefont {P.}~\bibnamefont {Ring}},\ }\bibfield
  {title} {\bibinfo {title} {Superdeformed rotational bands in the $a\sim
  140-150$ mass region: A cranked relativistic mean field description},\
  }\href@noop {} {\bibfield  {journal} {\bibinfo  {journal} {Nucl.\ Phys. A}\
  }\textbf {\bibinfo {volume} {608}},\ \bibinfo {pages} {107} (\bibinfo {year}
  {1996})}\BibitemShut {NoStop}%
\bibitem [{\citenamefont {Ring}\ \emph {et~al.}(1997)\citenamefont {Ring},
  \citenamefont {Gambhir},\ and\ \citenamefont {Lalalzissis}}]{RGL.97}%
  \BibitemOpen
  \bibfield  {author} {\bibinfo {author} {\bibfnamefont {P.}~\bibnamefont
  {Ring}}, \bibinfo {author} {\bibfnamefont {Y.~K.}\ \bibnamefont {Gambhir}},\
  and\ \bibinfo {author} {\bibfnamefont {G.~A.}\ \bibnamefont {Lalalzissis}},\
  }\bibfield  {title} {\bibinfo {title} {Computer program for the relativistic
  mean field description of the ground state properties of even-even axially
  deformed nuclei},\ }\href@noop {} {\bibfield  {journal} {\bibinfo  {journal}
  {Comp. Phys. Comm.}\ }\textbf {\bibinfo {volume} {105}},\ \bibinfo {pages}
  {77} (\bibinfo {year} {1997})}\BibitemShut {NoStop}%
\bibitem [{\citenamefont {Nik\v{s}i\'{c}}\ \emph {et~al.}(2014)\citenamefont
  {Nik\v{s}i\'{c}}, \citenamefont {Paar}, \citenamefont {Vretenar},\ and\
  \citenamefont {Ring}}]{DIRHB-code.14}%
  \BibitemOpen
  \bibfield  {author} {\bibinfo {author} {\bibfnamefont {T.}~\bibnamefont
  {Nik\v{s}i\'{c}}}, \bibinfo {author} {\bibfnamefont {N.}~\bibnamefont
  {Paar}}, \bibinfo {author} {\bibfnamefont {D.}~\bibnamefont {Vretenar}},\
  and\ \bibinfo {author} {\bibfnamefont {P.}~\bibnamefont {Ring}},\ }\bibfield
  {title} {\bibinfo {title} {Dirhb - a relativistic self-consistent mean-field
  framework for atomic nuclei},\ }\href
  {https://doi.org/10.1016/j.cpc.2014.02.027} {\bibfield  {journal} {\bibinfo
  {journal} {Comp. Phys. Comm.}\ }\textbf {\bibinfo {volume} {185}},\ \bibinfo
  {pages} {1808 } (\bibinfo {year} {2014})}\BibitemShut {NoStop}%
\bibitem [{\citenamefont {Osei}\ \emph
  {et~al.}(2025{\natexlab{b}})\citenamefont {Osei}, \citenamefont {Afanasjev},
  \citenamefont {Taninah}, \citenamefont {Dalbah}, \citenamefont {Perera},
  \citenamefont {Dzuba},\ and\ \citenamefont {Flambaum}}]{NL5Z-DDMEZ-PCZ}%
  \BibitemOpen
  \bibfield  {author} {\bibinfo {author} {\bibfnamefont {B.}~\bibnamefont
  {Osei}}, \bibinfo {author} {\bibfnamefont {A.~V.}\ \bibnamefont {Afanasjev}},
  \bibinfo {author} {\bibfnamefont {A.}~\bibnamefont {Taninah}}, \bibinfo
  {author} {\bibfnamefont {A.}~\bibnamefont {Dalbah}}, \bibinfo {author}
  {\bibfnamefont {U.~C.}\ \bibnamefont {Perera}}, \bibinfo {author}
  {\bibfnamefont {V.~A.}\ \bibnamefont {Dzuba}},\ and\ \bibinfo {author}
  {\bibfnamefont {V.~V.}\ \bibnamefont {Flambaum}},\ }\bibfield  {title}
  {\bibinfo {title} {Further steps toward the next generation of covariant
  energy density functionals},\ }\href@noop {} {\bibfield  {journal} {\bibinfo
  {journal} {Phys. Rev. C}\ }\textbf {\bibinfo {volume} {112}},\ \bibinfo
  {pages} {044304} (\bibinfo {year} {2025}{\natexlab{b}})}\BibitemShut
  {NoStop}%
\bibitem [{\citenamefont {P{\"o}schl}\ \emph {et~al.}(1997)\citenamefont
  {P{\"o}schl}, \citenamefont {Vretenar}, \citenamefont {Lalazissis},\ and\
  \citenamefont {Ring}}]{PVLR.97}%
  \BibitemOpen
  \bibfield  {author} {\bibinfo {author} {\bibfnamefont {W.}~\bibnamefont
  {P{\"o}schl}}, \bibinfo {author} {\bibfnamefont {D.}~\bibnamefont
  {Vretenar}}, \bibinfo {author} {\bibfnamefont {G.~A.}\ \bibnamefont
  {Lalazissis}},\ and\ \bibinfo {author} {\bibfnamefont {P.}~\bibnamefont
  {Ring}},\ }\bibfield  {title} {\bibinfo {title} {Relativistic
  hartree-bogoliuobov theory with finite range pairing forces in coordinate
  space: Neutron halo in light nuclei},\ }\href@noop {} {\bibfield  {journal}
  {\bibinfo  {journal} {Phys. Rev. Lett.}\ }\textbf {\bibinfo {volume} {79}},\
  \bibinfo {pages} {3841} (\bibinfo {year} {1997})}\BibitemShut {NoStop}%
\bibitem [{\citenamefont {Zhou}\ \emph {et~al.}(2003)\citenamefont {Zhou},
  \citenamefont {Meng},\ and\ \citenamefont {Ring}}]{ZMR.03}%
  \BibitemOpen
  \bibfield  {author} {\bibinfo {author} {\bibfnamefont {S.-G.}\ \bibnamefont
  {Zhou}}, \bibinfo {author} {\bibfnamefont {J.}~\bibnamefont {Meng}},\ and\
  \bibinfo {author} {\bibfnamefont {P.}~\bibnamefont {Ring}},\ }\bibfield
  {title} {\bibinfo {title} {Spherical relativistic hartree theory in a
  woods-saxon basis},\ }\href@noop {} {\bibfield  {journal} {\bibinfo
  {journal} {Phys. Rev. C}\ }\textbf {\bibinfo {volume} {68}},\ \bibinfo
  {pages} {034323} (\bibinfo {year} {2003})}\BibitemShut {NoStop}%
\bibitem [{\citenamefont {Tian}\ \emph {et~al.}(2009)\citenamefont {Tian},
  \citenamefont {Ma},\ and\ \citenamefont {Ring}}]{TMR.09}%
  \BibitemOpen
  \bibfield  {author} {\bibinfo {author} {\bibfnamefont {Y.}~\bibnamefont
  {Tian}}, \bibinfo {author} {\bibfnamefont {Z.~Y.}\ \bibnamefont {Ma}},\ and\
  \bibinfo {author} {\bibfnamefont {P.}~\bibnamefont {Ring}},\ }\bibfield
  {title} {\bibinfo {title} {A finite range pairing force for density
  functional theory in superfluid nuclei},\ }\href@noop {} {\bibfield
  {journal} {\bibinfo  {journal} {Phys.\ Lett. B}\ }\textbf {\bibinfo {volume}
  {676}},\ \bibinfo {pages} {44} (\bibinfo {year} {2009})}\BibitemShut
  {NoStop}%
\bibitem [{\citenamefont {Teeti}\ and\ \citenamefont
  {Afanasjev}(2021)}]{TA.21}%
  \BibitemOpen
  \bibfield  {author} {\bibinfo {author} {\bibfnamefont {S.}~\bibnamefont
  {Teeti}}\ and\ \bibinfo {author} {\bibfnamefont {A.~V.}\ \bibnamefont
  {Afanasjev}},\ }\bibfield  {title} {\bibinfo {title} {Global study of
  separable pairing interaction in covariant density functional theory},\
  }\href@noop {} {\bibfield  {journal} {\bibinfo  {journal} {Phys. Rev. C}\
  }\textbf {\bibinfo {volume} {103}},\ \bibinfo {pages} {034310} (\bibinfo
  {year} {2021})}\BibitemShut {NoStop}%
\bibitem [{\citenamefont {Perera}\ and\ \citenamefont
  {Afanasjev}(2022)}]{PA.22}%
  \BibitemOpen
  \bibfield  {author} {\bibinfo {author} {\bibfnamefont {U.~C.}\ \bibnamefont
  {Perera}}\ and\ \bibinfo {author} {\bibfnamefont {A.~V.}\ \bibnamefont
  {Afanasjev}},\ }\bibfield  {title} {\bibinfo {title} {Bubble nuclei:
  Single-particle versus coulomb interaction effects},\ }\href
  {https://doi.org/10.1103/PhysRevC.106.024321} {\bibfield  {journal} {\bibinfo
   {journal} {Phys. Rev. C}\ }\textbf {\bibinfo {volume} {106}},\ \bibinfo
  {pages} {024321} (\bibinfo {year} {2022})}\BibitemShut {NoStop}%
\bibitem [{\citenamefont {Perera}\ and\ \citenamefont
  {Afanasjev}(2023)}]{PA.23}%
  \BibitemOpen
  \bibfield  {author} {\bibinfo {author} {\bibfnamefont {U.~C.}\ \bibnamefont
  {Perera}}\ and\ \bibinfo {author} {\bibfnamefont {A.~V.}\ \bibnamefont
  {Afanasjev}},\ }\bibfield  {title} {\bibinfo {title} {Differential charge
  radii: Proton-neutron interaction effects},\ }\href
  {https://doi.org/10.1103/PhysRevC.107.064321} {\bibfield  {journal} {\bibinfo
   {journal} {Phys. Rev. C}\ }\textbf {\bibinfo {volume} {107}},\ \bibinfo
  {pages} {064321} (\bibinfo {year} {2023})}\BibitemShut {NoStop}%
\bibitem [{\citenamefont {Zhao}\ \emph {et~al.}(2010)\citenamefont {Zhao},
  \citenamefont {Li}, \citenamefont {Yao},\ and\ \citenamefont
  {Meng}}]{PC-PK1}%
  \BibitemOpen
  \bibfield  {author} {\bibinfo {author} {\bibfnamefont {P.~W.}\ \bibnamefont
  {Zhao}}, \bibinfo {author} {\bibfnamefont {Z.~P.}\ \bibnamefont {Li}},
  \bibinfo {author} {\bibfnamefont {J.~M.}\ \bibnamefont {Yao}},\ and\ \bibinfo
  {author} {\bibfnamefont {J.}~\bibnamefont {Meng}},\ }\bibfield  {title}
  {\bibinfo {title} {New parametrization for the nuclear covariant energy
  density functional with a point-coupling interaction},\ }\href@noop {}
  {\bibfield  {journal} {\bibinfo  {journal} {Phys.\ Rev. C}\ }\textbf
  {\bibinfo {volume} {82}},\ \bibinfo {pages} {054319} (\bibinfo {year}
  {2010})}\BibitemShut {NoStop}%
\bibitem [{\citenamefont {Nik\v{s}i\'{c}}\ \emph {et~al.}(2008)\citenamefont
  {Nik\v{s}i\'{c}}, \citenamefont {Vretenar},\ and\ \citenamefont
  {Ring}}]{DD-PC1}%
  \BibitemOpen
  \bibfield  {author} {\bibinfo {author} {\bibfnamefont {T.}~\bibnamefont
  {Nik\v{s}i\'{c}}}, \bibinfo {author} {\bibfnamefont {D.}~\bibnamefont
  {Vretenar}},\ and\ \bibinfo {author} {\bibfnamefont {P.}~\bibnamefont
  {Ring}},\ }\bibfield  {title} {\bibinfo {title} {Relativistic nuclear energy
  density functionals: adjusting parameters to binding energies},\ }\href@noop
  {} {\bibfield  {journal} {\bibinfo  {journal} {Phys.\ Rev. C}\ }\textbf
  {\bibinfo {volume} {78}},\ \bibinfo {pages} {034318} (\bibinfo {year}
  {2008})}\BibitemShut {NoStop}%
\bibitem [{\citenamefont {Angeli}\ and\ \citenamefont
  {Marinova}(2013)}]{AM.13}%
  \BibitemOpen
  \bibfield  {author} {\bibinfo {author} {\bibfnamefont {I.}~\bibnamefont
  {Angeli}}\ and\ \bibinfo {author} {\bibfnamefont {K.~P.}\ \bibnamefont
  {Marinova}},\ }\bibfield  {title} {\bibinfo {title} {Table of experimental
  nuclear ground state charge radii: An update},\ }\href@noop {} {\bibfield
  {journal} {\bibinfo  {journal} {At.\ Data Nucl.\ Data Tables}\ }\textbf
  {\bibinfo {volume} {99}},\ \bibinfo {pages} {69} (\bibinfo {year}
  {2013})}\BibitemShut {NoStop}%
\bibitem [{\citenamefont {Shanks}(1955)}]{Shanks.55}%
  \BibitemOpen
  \bibfield  {author} {\bibinfo {author} {\bibfnamefont {D.}~\bibnamefont
  {Shanks}},\ }\bibfield  {title} {\bibinfo {title} {Non-linear transformation
  of divergent and slowly convergent sequences},\ }\href@noop {} {\bibfield
  {journal} {\bibinfo  {journal} {J. of Math. and Phys.}\ }\textbf {\bibinfo
  {volume} {34}},\ \bibinfo {pages} {1} (\bibinfo {year} {1955})}\BibitemShut
  {NoStop}%
\bibitem [{\citenamefont {Wynn}(1956)}]{Wynn.56}%
  \BibitemOpen
  \bibfield  {author} {\bibinfo {author} {\bibfnamefont {P.}~\bibnamefont
  {Wynn}},\ }\bibfield  {title} {\bibinfo {title} {On a device for computing
  the $e_m(s_n)$ transformation},\ }\href@noop {} {\bibfield  {journal}
  {\bibinfo  {journal} {Mathematical Tables and Other Aids to Computation}\
  }\textbf {\bibinfo {volume} {190}},\ \bibinfo {pages} {91} (\bibinfo {year}
  {1956})}\BibitemShut {NoStop}%
\bibitem [{\citenamefont {Roache}\ and\ \citenamefont {Knupp}(1993)}]{RK.93}%
  \BibitemOpen
  \bibfield  {author} {\bibinfo {author} {\bibfnamefont {P.~J.}\ \bibnamefont
  {Roache}}\ and\ \bibinfo {author} {\bibfnamefont {P.~M.}\ \bibnamefont
  {Knupp}},\ }\bibfield  {title} {\bibinfo {title} {Completed richardson
  extrapolation},\ }\href@noop {} {\bibfield  {journal} {\bibinfo  {journal}
  {Comm. in Num. Meth. in Eng.}\ }\textbf {\bibinfo {volume} {9}},\ \bibinfo
  {pages} {365} (\bibinfo {year} {1993})}\BibitemShut {NoStop}%
\bibitem [{\citenamefont {Afanasjev}\ \emph {et~al.}(2026)\citenamefont
  {Afanasjev}, \citenamefont {Litvinova},\ and\ \citenamefont {Osei}}]{ALO.26}%
  \BibitemOpen
  \bibfield  {author} {\bibinfo {author} {\bibfnamefont {A.~V.}\ \bibnamefont
  {Afanasjev}}, \bibinfo {author} {\bibfnamefont {E.}~\bibnamefont
  {Litvinova}},\ and\ \bibinfo {author} {\bibfnamefont {B.}~\bibnamefont
  {Osei}},\ }\bibfield  {title} {\bibinfo {title} {Basis truncation,
  statistical errors, and systematic uncertainties in relativistic approaches
  to nuclear response},\ }\href@noop {} {\bibfield  {journal} {\bibinfo
  {journal} {submitted to Phys. Rev. C, see also arXiv:2512.23125}\ } (\bibinfo
  {year} {2026})}\BibitemShut {NoStop}%
\bibitem [{\citenamefont {Litvinova}\ and\ \citenamefont
  {Afanasjev}(2011)}]{LA.11}%
  \BibitemOpen
  \bibfield  {author} {\bibinfo {author} {\bibfnamefont {E.~V.}\ \bibnamefont
  {Litvinova}}\ and\ \bibinfo {author} {\bibfnamefont {A.~V.}\ \bibnamefont
  {Afanasjev}},\ }\bibfield  {title} {\bibinfo {title} {Dynamics of nuclear
  single-particle structure in covariant theory of particle-vibration coupling:
  From light to superheavy nuclei},\ }\href@noop {} {\bibfield  {journal}
  {\bibinfo  {journal} {Phys.\ Rev. C}\ }\textbf {\bibinfo {volume} {84}},\
  \bibinfo {pages} {014305} (\bibinfo {year} {2011})}\BibitemShut {NoStop}%
\bibitem [{\citenamefont {Afanasjev}\ and\ \citenamefont
  {Abusara}(2018)}]{AA.18}%
  \BibitemOpen
  \bibfield  {author} {\bibinfo {author} {\bibfnamefont {A.~V.}\ \bibnamefont
  {Afanasjev}}\ and\ \bibinfo {author} {\bibfnamefont {H.}~\bibnamefont
  {Abusara}},\ }\bibfield  {title} {\bibinfo {title} {From cluster structures
  to nuclear molecules: The role of nodal structure of the single-particle wave
  functions},\ }\href@noop {} {\bibfield  {journal} {\bibinfo  {journal} {Phys.
  Rev. C}\ }\textbf {\bibinfo {volume} {97}},\ \bibinfo {pages} {024329}
  (\bibinfo {year} {2018})}\BibitemShut {NoStop}%
\bibitem [{\citenamefont {Baran}\ \emph {et~al.}(2015)\citenamefont {Baran},
  \citenamefont {Kowal}, \citenamefont {Reinhard}, \citenamefont {Robledo},
  \citenamefont {Staszczak},\ and\ \citenamefont {Warda}}]{BKRRSW.15}%
  \BibitemOpen
  \bibfield  {author} {\bibinfo {author} {\bibfnamefont {A.}~\bibnamefont
  {Baran}}, \bibinfo {author} {\bibfnamefont {M.}~\bibnamefont {Kowal}},
  \bibinfo {author} {\bibfnamefont {P.-G.}\ \bibnamefont {Reinhard}}, \bibinfo
  {author} {\bibfnamefont {L.~M.}\ \bibnamefont {Robledo}}, \bibinfo {author}
  {\bibfnamefont {A.}~\bibnamefont {Staszczak}},\ and\ \bibinfo {author}
  {\bibfnamefont {M.}~\bibnamefont {Warda}},\ }\bibfield  {title} {\bibinfo
  {title} {Fission barriers and probabilities of spontaneous fission for
  elements with z$\geq$ 100},\ }\href@noop {} {\bibfield  {journal} {\bibinfo
  {journal} {Nucl.\ Phys. A}\ }\textbf {\bibinfo {volume} {944}},\ \bibinfo
  {pages} {442} (\bibinfo {year} {2015})}\BibitemShut {NoStop}%
\bibitem [{\citenamefont {Schunck}\ and\ \citenamefont
  {Robledo}(2016)}]{SR.16}%
  \BibitemOpen
  \bibfield  {author} {\bibinfo {author} {\bibfnamefont {N.}~\bibnamefont
  {Schunck}}\ and\ \bibinfo {author} {\bibfnamefont {L.~M.}\ \bibnamefont
  {Robledo}},\ }\bibfield  {title} {\bibinfo {title} {Microscopic theory of
  nuclear fission: A review},\ }\href@noop {} {\bibfield  {journal} {\bibinfo
  {journal} {Rep. Prog. Phys.}\ }\textbf {\bibinfo {volume} {79}},\ \bibinfo
  {pages} {116301} (\bibinfo {year} {2016})}\BibitemShut {NoStop}%
\bibitem [{\citenamefont {Taninah}\ \emph {et~al.}(2020)\citenamefont
  {Taninah}, \citenamefont {Agbemava},\ and\ \citenamefont
  {Afanasjev}}]{TAA.20}%
  \BibitemOpen
  \bibfield  {author} {\bibinfo {author} {\bibfnamefont {A.}~\bibnamefont
  {Taninah}}, \bibinfo {author} {\bibfnamefont {S.~E.}\ \bibnamefont
  {Agbemava}},\ and\ \bibinfo {author} {\bibfnamefont {A.~V.}\ \bibnamefont
  {Afanasjev}},\ }\bibfield  {title} {\bibinfo {title} {Covariant density
  functional theory input for $r$-process simulations in actinides and
  superheavy nuclei: The ground state and fission properties},\ }\href@noop {}
  {\bibfield  {journal} {\bibinfo  {journal} {Phys. Rev. C}\ }\textbf {\bibinfo
  {volume} {102}},\ \bibinfo {pages} {054330} (\bibinfo {year}
  {2020})}\BibitemShut {NoStop}%
\end{thebibliography}%

\end{document}